\documentclass[aps,pre,twocolumn,floatfix,nobalancelastpage]{revtex4-1}
\usepackage{makeidx}
\usepackage{bm}
\usepackage{amssymb}
\usepackage{amssymb,amsmath}
\usepackage{graphicx}
\usepackage{graphics}
\usepackage{epstopdf}
\usepackage{dsfont}
\usepackage{array}
\usepackage{slashed}
\usepackage{bm}
\usepackage{verbatim}

\newcommand{\nn}{\nonumber}

\usepackage{color}

\newcommand{\rev}{}

\begin{document}
\title{On the disorder-driven quantum transition in three-dimensional relativistic metals}

\author{T. Louvet, D. Carpentier, and A. A. Fedorenko}
\affiliation{\mbox{Univ Lyon, ENS de Lyon, Univ Claude Bernard, CNRS, Laboratoire de Physique, F-69342 Lyon, France}}

\date{December 23, 2016}

\begin{abstract}

The  Weyl semimetals  are topologically protected from a gap opening against weak disorder in three dimensions.
However, a strong disorder  drives this relativistic semimetal through  a quantum transition towards a diffusive metallic phase characterized by
a finite density of states at the band crossing. This transition is usually described by a perturbative renormalization group in  $d=2+\varepsilon$ of a $U(N)$ Gross-Neveu model in the limit $N \to 0$. Unfortunately,  this model is not multiplicatively renormalizable
in $2+\varepsilon$ dimensions:  An infinite number of relevant operators are required to describe the critical behavior.
Hence its use in a quantitative description of the transition  beyond one-loop is at least questionable.
{\rev We propose an alternative route, building on the correspondence between the Gross-Neveu and Gross-Neveu-Yukawa models developed  in the context of high energy physics. It results in a model of Weyl fermions
with a random non-Gaussian  imaginary potential which allows one to study the critical properties of the transition
within a $d=4-\varepsilon$ expansion. We also discuss the characterization of the transition by the multifractal spectrum
of wave functions. }
\end{abstract}

\maketitle

\textit{Introduction. -}  After the discovery of graphene,  materials with a relativistic-like
spectrum of electronic excitations have become a popular subject which currently
drives several hot topics in condensed matter physics.
Examples include three dimensional materials such as $Na_3Bi$ and $Cd_3As_2$
which have been identified as Dirac semimetals~\cite{Liu:2014,Neupane:2014,Borisenko:2014}.
The twofold band degeneracy of Dirac semimetals can be lifted by
breaking time or inversion symmetry as it happens in  $TaAs$  and  $NbAs$
leading to the so called Weyl semimetal~\cite{Xu:2015a,Xu:2015b}.
The latter is topologically protected from a gap opening against small
perturbations. Indeed, real materials inevitably contain disorder of different kinds, which turn out to be irrelevant in
the renormalization group (RG) sense. The weakly disordered materials remain in a semimetallic 
phase~\cite{Goswami:2011,Hosur:2012,Ominato:2014}:
At the nodal point, the system is characterized by a density of states (DOS) vanishing quadratically with energy
up to exponentially small corrections due to rare events~\cite{Nandkishore:2014,Pixley:2016}.
It exhibits a vanishing zero-frequency optical conductivity~\cite{Roy:2016b}
and a pseudoballistic transport~\cite{Sbierski:2014}.
However, as was pointed for the first time in Refs.~\cite{Fradkin:1986}, 
a  strong enough disorder
may drive the system into a diffusive phase with a finite DOS, optical conductivity
and diffusive transport at zero energy.
The semimetal to diffusive metal transition has been numerically studied for several models~\cite{Kobayashi:2014,Sbierski:2015,Chen:2015}
including both  the Dirac and Weyl semimetals.
For the simplest scalar potential considered in this 
Rapid Communication all of them belong to the same
universality class. However, the nature of the disordered phase and its protection against Anderson localization
depends on the precise nature of the phase, {\it e.g.} Dirac versus
Weyl semimetals~\cite{Altland:2015,Altland:2016,Syzranov:2015,Syzranov:2015c,Garttner:2015}.

It is now believed that the disorder-driven  quantum transition from a single-cone Weyl semimetal to a diffusive phase
is related to the chiral transition well studied in high energy physics and described by the 3D $U(N)$ Gross-Neveu (GN) model,
but in the unusual limit of a vanishing number of components $N\to 0$.
This relation has been confirmed by direct calculations to two-loop order on the initial Weyl model using either supersymmetry~\cite{Syzranov:2015b}  or
replica methods~\cite{Roy:2014,Roy:2016}.
{\rev The massless GN model possesses a chiral symmetry which is spontaneously broken for sufficiently strong interactions.
For the disordered Weyl fermions this transition translates into the appearance of a finite DOS at the nodal point for disorders stronger than a critical value.}

However, we recall here that the $U(N)$ GN model is not multiplicatively renormalizable in dimension $d=2+\epsilon_2 >2$ :
This manifests itself in the generation of an infinite number of relevant operators along the RG flow beyond two-loop order.
Moreover, these relevant operators collapse into
a few operators when extrapolating this technique to $d=3$.
 This casts some doubts about the direct applicability of  this approach to the Weyl fermion problem in $d=3$.
Taking into account the inherent difficulties of this $d=2+\varepsilon_2$ expansion we propose a different approach based on a  $d=4-\varepsilon_4$ expansion
to study the disorder-driven transition in the Weyl semimetals.
In this approach we build on the known correspondence between the GN model and the $U(N)$ Gross-Neveu-Yukawa (GNY) model for $2\le d\le4$~\cite{Zinn-Justin:1986},
which is similar to the relation of the $O(N)$ non-linear $\sigma$-model with respect to the $O(N)$ $\varphi^4$ model~\cite{Hasenfratz:1991}.
Besides the fermionic field, the GNY model involves an additional scalar bosonic field. {\rev In the limit of $N \to 0$
it can be interpreted as a random non-Gaussian imaginary potential.}
The equivalence between this and the initial problem sheds light on the quantum transition
and we discuss several of its possible  consequences.
For instance, it  allows one to calculate the critical exponents in a systematic  controllable way, since this model is
renormalizable in $4-\varepsilon_4$ dimensions.

\textit{Model. -}
The action of the $d$-dimensional
relativistic fermions moving in {\rev the random disorder potential
$V(\mathbf{r})$   can be written as~\cite{Ludwig:1994}
\begin{equation}
\label{eq:action0}
S_{\mathrm{\tiny Weyl}} =
	\int d^d r \int d \omega~ \bar{\psi}(\mathbf{r},-\omega)\left[ - i  \slashed{\partial}  - i \omega     + V(\mathbf{r})  \right]   \psi(\mathbf{r},\omega) ,
\end{equation}
where $ \slashed{\partial}= \gamma_{\mu} \partial^{\mu}$  and $\omega$ is a Matsubara frequency}.
The  $\gamma_i$ are elements of the Clifford algebra
which satisfy the anticommutation relations:
$\gamma_i \gamma_j + \gamma_j \gamma_i = 2 \delta_{ij} \mathbb{I}$, and $i,j=1,...,d$.
The Weyl fermions corresponds to $d=3$  and $\gamma_i = \sigma_i$ given by the Pauli matrices.
{\rev To average over disorder distribution $P_V[V]$ we introduce $N$ copies of the
system  so that  physical observables  can be calculated in the limit of  $N\to 0$
using the replicated action
\begin{eqnarray}
\int \prod_{\alpha=1}^N \mathcal{D}\{\psi_\alpha\} e^{-S_\mathrm{repl}}
= \int \mathcal{D}V   P_V[V] \prod_{\alpha=1}^N \mathcal{D}\{\psi_\alpha\} e^{-S_{\mathrm{\tiny Weyl}}^\alpha},  \nn
\end{eqnarray}
where $\mathcal{D}\{\psi_\alpha\} =   \mathcal{D} {\bar \psi_\alpha} \mathcal{D}{\psi_\alpha}$.
We neglect the possible presence of long-range spatial correlations
which can modify the critical properties~\cite{Fedorenko:2012} and take the distribution of disorder potential to be Gaussian,  $P_V[V]\sim e^{-\frac1{2 \Delta_0} \int d^d r V({\bf r})^2}$. This yields
\begin{multline} \label{eq:action1}
 S_{\mathrm{repl}} =
	\int d^d r \int d \omega  \left [ - i  \bar{\psi}_{\alpha}(\mathbf{r},-\omega)( \slashed{\partial} + \omega  ) \right.
	\psi_{\alpha}(\mathbf{r},\omega)
 \\
 \left. - \frac{\Delta_0}2
 	 \int d\omega'  \bar{\psi}_{\alpha}(\mathbf{r},-\omega) \psi_{\alpha}(\mathbf{r},\omega )
	\bar{\psi}_{\beta}(\mathbf{r},-\omega') \psi_{\beta}(\mathbf{r}, \omega')\right ],
\end{multline}
where a summation over $\alpha,\beta$ is implied and disorder generates an attractive interaction
between different replicas.
It turns out that  the Green's functions  computed for the action \eqref{eq:action1} at fixed energy $\omega$ in the limit $N\to 0$
can be deduced from the  $d$-dimensional $U(N)$ GN model
\begin{equation} \label{eq:action-naiveGN}
  S_{\mathrm{GN}} =  - \int d^d r  \left[  {\bar {\bm \chi}} \cdot ( \slashed{\partial} + \omega  ) {\bm \chi}
  - \frac{\Delta_0}2 ({\bar {\bm \chi}}{\bm \chi})({\bar {\bm \chi}}{\bm \chi}) \right],
\end{equation}
which appears here with a negative (attractive) coupling constant in terms of new fields  $\bar{\bm \chi} = i \bar{\bm \psi} ({\bm r},-\omega)$  and ${\bm \chi} = { \bm \psi} ({\bm r},\omega)$~\cite{Syzranov:2015b}.
}

\textit{$2+\varepsilon_2$ expansion. -}
{ \rev  We now show that a renormalization procedure based on the model~\eqref{eq:action-naiveGN} is inherently flawed beyond the
 two-loop order of previous studies~\cite{Schuessler:2009,Roy:2014,Syzranov:2015b}: 
 The problem is related to the extension of the
Clifford algebra to arbitrary dimensions necessary within the renormalization scheme.
Indeed, in $2< d=2+\varepsilon_2 <3$, the product $\gamma_i\gamma_j$ cannot be expressed as a linear
combination of $\gamma_i$ so that the Clifford algebra becomes infinite-dimensional.
It is then convenient to use antisymmetrized  products such as
 $\gamma^{(n)}_{\vec{A}}=\textrm{As}[\gamma_{a_1}...\gamma_{a_n}]$,
where we have introduced  the notation $\vec{A}=\{ a_1, ...  , a_n \}$, as a basis in this infinite-dimensional space
so that one does not need any explicit representation of these objects to perform calculations.
Thus,  along the RG flow an infinite number of corresponding operators are generated,
of the form
$V^{(n)} =	 \left(\bar{\chi}_\alpha\gamma^{(n)}_{\vec{A}} \chi_\alpha \right)\cdot \left(\bar{\chi}_\beta\gamma^{(n)}_{\vec{A}} \chi_\beta \right)$,
where a summation over $\alpha$, $\beta$ and $\vec{A}$ is implied.
The minimal multiplicatively renormalizable model replacing \eqref{eq:action-naiveGN} hence reads } 
\begin{eqnarray} \label{eq:action-GN}
  S_{\mathrm{GN}} &=&  - \int d^d r  \left[  {\bar {\bm \chi}} \cdot ( \slashed{\partial} + \omega  ) {\bm \chi}
  - \frac{1}2 \sum\limits_{n=0}^{\infty} {\Delta_n}  V^{(n)} \right].
\end{eqnarray}
{\rev As an example, let us consider} the three-loop order for which only the operators $V^{(3)}$ and $V^{(4)}$ are generated \cite{Vasilev:1997}.
The corresponding RG flow equations are given
in the limit $N \to 0$ by the $\beta$-functions,
\begin{subequations}
\begin{align}
& \frac{\partial \Delta_0 }{\partial \ln L }  = - \varepsilon_2 \Delta_0  + 4 \Delta_0^2 +  8 \Delta_0^3 + 28 \Delta_0^4, \label{eq-beta1} \\
& \frac{\partial \Delta_3 }{\partial \ln L }  = -\varepsilon_2 \Delta_3 + a  \Delta_0^4 +16 \Delta_0 \Delta_4 + 8 \Delta_0\Delta_3.  \label{eq-beta2} \\
& \frac{\partial \Delta_4 }{\partial \ln L }  = -\varepsilon_2 \Delta_4 -4 \Delta_0\Delta_3 -12 \Delta_0 \Delta_4, \label{eq-beta3}
\end{align}
\end{subequations}
where $a= -4 + \zeta(3)$ and $\zeta(x)$ is the Riemann zeta function.
To this order, the fixed point (FP) describing the transition reads
$\Delta_0^*=\varepsilon_2 /4 - \varepsilon_2 ^2/8 + \varepsilon_2 ^3 / 64 + O(\varepsilon_2^4),
\Delta_3^*=a \varepsilon_2 ^2 / 96 - 23 a \varepsilon_2 ^3/1152+O(\varepsilon_2 ^4),
\Delta_4^*= - a \varepsilon_2 ^2 / 384 + 49 a \varepsilon_2 ^3 / 9216 +O (\varepsilon_2 ^4).
$
Note the peculiarity of the limit $N \to 0$ where, while $\Delta_0^*$ is of order $\varepsilon_2$,  the  generated operators
are order of $\varepsilon_{2}^2$ instead of $\varepsilon_{2}^3$ expected in the three-loop order.
{\rev The critical exponent of the correlation length divergence at the transition as $\xi~\sim |\Delta-\Delta^*|^{-\nu} $
reads
$ 1/\nu  = \varepsilon_2 +\frac12 \varepsilon_2 ^2 +\frac38  \varepsilon_2 ^3 +O\left(\varepsilon_2^4\right)$.
For a Weyl semimetal ($\varepsilon_4 = 1$) we find: $\nu = 0.533$ (direct substitution)\footnote{
Other resummation methods give
$\nu = 0.333$ by Pad\'{e} [2/1] and $\nu = 0.375$
by Pad\'{e} [1/2]. Note that the Pad\'{e}-Borel[2/1] has a pole, but the principal value integral gives $\nu=0.57$.}.
}
Crucially,  the validity of this renormalization picture directly in dimension $d=3$ is questionable: The Clifford algebra is then of finite dimension.
Hence all the operators $V^{(n)}$ generated  by the RG flow beyond three-loops
either disappear (evanescent operators) or  collapse on a few operators when extending $d=2+\varepsilon_{2}\to 3$.
Contrary to the two-dimensional case~\cite{Gracey:2008}
no standard  projecting procedure exists to reduce the $\beta$-functions for these
evanescent  operators to the  $\beta$-function for the remaining operators  in $d=3$.

\textit{$4-\varepsilon_4$ expansion. -}
{\rev  Here we propose another way to describe the quantum transition  alternative to the use of~\eqref{eq:action-GN}.
This new approach is of interest beyond the quantitative
calculations since it provides an example of  a physical model possessing the same quantum critical properties as
the disordered Weyl fermions.
It is based on the well known correspondence between the critical properties of the $U(N)$ GN and GNY models~\cite{Zinn-Justin:1986}
which we transpose in the context of the disordered relativistic fermions associated with the $N\to 0$ limit.
Contrary to the GN model the GNY model is renormalizable in dimension $d=4-\varepsilon_4$:
 critical properties of the transition can be obtained to any order without  generating an infinite number of relevant operators.}
In the $U(N)$ GNY model, an additional scalar field $\phi$ is introduced, and the action reads
\begin{multline}
S_{GNY} =  \int d^d r  \biggl[ - {\bar \chi}_\alpha ( \slashed{\partial} + \sqrt{g} \phi ) \chi_\alpha   \\
+ \frac12 (\nabla \phi)^2 + \frac{\mu}{2}  \phi^2 + \frac{\lambda}{4\text{!}} \phi^4 \biggr].
\label{eq:action-GNY}
\end{multline}
 {\rev
  In terms of the initial fields $\bar{\bm  \psi} = - i \bar {\bm \chi} , {\bm \chi} = {\bm \psi}$,
 the GNY model~\eqref{eq:action-GNY}  corresponds to the Weyl fermions at $\omega =0$
 coupled to an imaginary random potential \footnote{Similarly to going from \eqref{eq:action1} to \eqref{eq:action-naiveGN},
 the coupling between fermions of different  frequencies is irrelevant for constant $\omega$ properties in the limit $N\to 0$.}
\begin{equation}
S^\alpha =  \int d^d r ~ {\bar \psi}_\alpha ( -i \slashed{\partial} - i \sqrt{g} \phi ) \psi_\alpha,
\label{eq:action-img}
\end{equation}
 with the random potential distribution given by
\begin{equation}
\label{eq:dist-phi}
 P[\phi] \propto \mathrm{exp} \left(- \int d^dr~ \biggl[ \frac12 (\nabla \phi)^2 + \frac{\mu}{2}  \phi^2 + \frac{\lambda}{4\text{!}} \phi^4 \biggr] \right).
\end{equation}
Such a random imaginary potential is unusual: It can be interpreted as an effective inverse life-time (imaginary part of a self-energy), which appear to be randomly distributed. The transposition of the GN - GNY correspondence in the
context of disordered Weyl fermions amounts to the equivalence between  a random Gaussian scalar potential and a non-Gaussian imaginary  field distributed  according to \eqref{eq:dist-phi}.
Studying the relevance of this correspondence beyond these simple distribution functions will be of great interest.
}

\begin{figure}
\includegraphics[width=83mm]{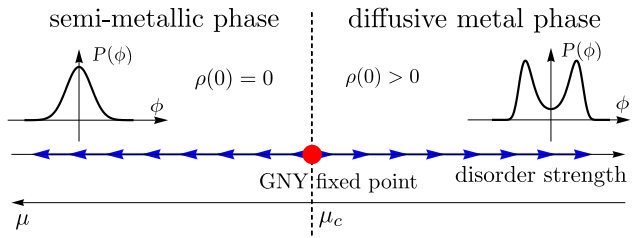}
\caption{
  \label{fig:3DFlow} \rev
Schematic projection of the RG flow for the $U(N)$ GNY model~(\ref{eq:action-GNY}) in the three-parameter space: $g$, $\lambda$ and $\mu$ onto an unstable direction along $\mu$. On the left side from the GNY FP, the  flow towards large $\mu$  corresponds to the semi-metallic phase with vanishing DOS at the band crossing and a Gaussian distribution of  field $\phi$. On the right side
 the flow towards small $\mu$ drives the system towards a diffusive metal with a finite DOS and non-Gaussian distribution of field $\phi$. }
\end{figure}

{\rev The transition  within the GNY model can be understood at the mean-field level:
 (i)~For $\mu >0$, the typical (most probable) value of the scalar field $\phi$ vanishes
 and we recover a theory of free fermions. This corresponds to a phase where the disorder potential $\sqrt{g} \phi$ is Gaussian, distributed around $\phi=0$;
 (ii)~on the other hand, for $\mu <0$,
the scalar field acquires a finite typical value. This translates into a finite density of states of the Weyl fermions at zero energy, $\rho(0)>0$.  In this phase, the mean-field distribution of the disorder potential $P[\phi]$ is peaked around opposite values (see Fig.~\ref{fig:3DFlow}) and the distribution is no longer Gaussian.
In the context of the high energy physics the generation of a finite typical value $\phi$
corresponds to breaking the chiral/spatial parity symmetry (in even/odd dimensions)
by generating a fermionic mass~\cite{Zinn-Justin:1986}.

Now let us discuss the critical properties of the transition in more details through a renormalization group analysis of the GNY model \eqref{eq:action-GNY}.
}
The correspondence between the critical properties of the GN and GNY models has been shown using  $1/N$ expansion and numerically
for finite $N$~\cite{Karkkainen:1994}.
Since the GNY model shows analytical behavior with $N$, we extend this correspondence between the two models in the limit $N\to 0$.
To renormalize the model~(\ref{eq:action-GNY}) we use a minimal subtraction scheme with dimensional regularization.
Introducing the momentum scale $\Lambda$, we define the dimensionless parameters:
${\tilde g} = \Lambda^{-\varepsilon_4} g,~  {\tilde \lambda} = \Lambda^{-\varepsilon_4} \lambda$, ${\tilde \mu} = \Lambda^{-2} \mu$.
Whereas the couplings $\lambda, g$ are multiplicatively renormalized, the parameter $\mu$ driving the transition acquires a
non-universal shift: One has to consider the flow of the deviation from the critical value, $\delta \mu = \mu - \mu_c$.
The RG flow equations read:
\begin{subequations}
\begin{align}
& \frac{\partial {\tilde \lambda}}{\partial \ln L}  = \varepsilon_4 {\tilde \lambda} - 3 {\tilde \lambda}^2 + \frac{17}{3}{\tilde \lambda}^3  , \\
& \frac{\partial {\tilde g}}{\partial \ln L}  = \varepsilon_4 {\tilde g}  - 6 {\tilde g^2} + \frac{9}{2} {\tilde g^3} + 4 {\tilde \lambda}  {\tilde g^2} - \frac16 {\tilde \lambda^2} {\tilde g}.
\end{align}
\end{subequations}
The critical fixed point is defined by
$g^{*} =  \frac16 \varepsilon_4   + \frac{71}{1296} \varepsilon_4^2 + O(\varepsilon_4^3)$,
$\lambda^{*} =  \frac13 \varepsilon_4   + \frac{17}{81} \varepsilon_4^2  + O(\varepsilon_4^3)$ and $\mu = \mu_c$.
The FP is IR stable in the directions $\lambda$ and $g$. $\mu$ is the only relevant variable; around the fixed point its scaling with the correlation length $\xi$
defines the critical length exponent~$\nu$: $|\delta \mu| \sim \xi^{-1/\nu}$.
We find to two-loop order~\cite{Supplementary}:
\begin{equation}
 \frac1{\nu}  =  2 - \frac{\varepsilon_4}3 -\frac{19}{162}\varepsilon_4^2 + O(\varepsilon_4^3).
 \end{equation}
The numerical value of the exponent $\nu$ to two-loop order
is given by $\nu=0.65$ (direct substitution $\varepsilon_4=1$)
\footnote{Resummation methods give
$\nu=0.67$ (Pad\'{e} [1/1]), $\nu = 0.699$ (Pad\'{e}-Borel [1/1]).}.

\textit{Discussion. - }
We studied the disorder driven transition of the 3D Weyl semimetals towards a diffusive metal.
We showed that the description of this transition using the $U(N)$ GN model in $2+\varepsilon_2$ dimensions
in the limit of $N \to 0$ encounters significant difficulties already
beyond one-loop approximations.  They  are related to the multiplicative non-renormalizability of the model and generation of
an infinite number of vertices whose three-loop corrections unexpectedly shift the fixed point to the order $\varepsilon_2^2$.
We have proposed an alternative approach based on the correspondence between the $U(N)$ GNY and GN models.
{\rev The previous numerical and analytical studies give values of the correlation length exponent
$\nu$ which lie in a broad range from $0.6$ to $1.5$~\cite{Kobayashi:2014,Sbierski:2015,Goswami:2011,Syzranov:2015b,Roy:2014,Roy:2016},
that can be related to the existence of a large number of relevant operators in the GN model.  The GNY model has only one relevant operator
and we find $\nu \approx 0.65-0.67$.
Beyond offering a well defined  framework
for an unambiguous description of the critical properties it relates the quantum transition of disordered Weyl fermions with
chemical potential fluctuations to  that of a model with spatially correlated and non Gaussian imaginary disorder.
We are confident that this novel correspondence   between two models of identical fermions  with distinct disorder potentials
opens interesting perspectives for further investigations such as functional renormalization group studies of this transition~\cite{Carpentier:2000}.
}

Let us discuss these results in view of recent work on the relevance of rare
disorder realizations around the transition~\cite{Nandkishore:2014,Pixley:2016}.
{\rev This is an important issue, since  the semimetal phase can be destabilized
not only by finite doping but also by the zero energy states emergent
from these rare disorder configurations.}
 The stability of a disordered fixed point with respect to fluctuations effects is related to the extended Harris  criterion
$\nu_\mathrm{FS}>2/d$ \cite{Chayes:1986} for the correlation length critical exponent.
The values of $\nu$ for both the GN and GNY models  violate this criterion at the order considered.
{
However, this inequality has to be satisfied by the finite size correlation exponent,
while there is no restriction on the intrinsic exponent usually probed by the RG methods: In principle, it can be different from the
first one~\cite{Pazmandi:1997}.

 On the other hand,  the relevance of rare fluctuations around the transition can manifest itself in the RG context by the development of
a strong deviation from the Gaussian distribution of disorder.  The corresponding cumulants are related to the composite operators
$\mathcal{O}_q= (\bar{\psi}_\alpha \psi_\alpha )^q$.  We find to order one-loop in the GN model
 the scaling dimension  of these operators $[\mathcal{O}_q] = (d-1)q - 2 q^2 \Delta_0^* +O(\Delta_0^{*2}) $.
 Thus, these operators with $q\ge4$ become naively relevant  {\rev at the FP of  (\ref{eq-beta1}) }
  for $\varepsilon_2>2/5$.  This observation
 suggests that strong deviations from the Gaussian distribution of disorder develop in $d=3$ ($\varepsilon_{2}=1$), which could explain the importance
 of rare disorder realizations.
 Indeed, in Refs.~\cite{Nandkishore:2014,Pixley:2016} it was shown  that the
 \emph{average} DOS at zero energy can be finite in  the semi-metallic phase due to contributions from rare events
  that leads to an avoided quantum transition.
For the  GNY model we also find~\cite{Supplementary} instanton-like solutions  similar to those observed in the GN model \cite{Nandkishore:2014,Pixley:2016} and which are responsible for the contribution of the rare events to the DOS.
Whether such instanton solutions can be accounted for by a more refined renormalization of the
distribution of disorder beyond the GNY model remains a question of interest.

Let us note, however, that an alternative characterization of the transition exists, less sensitive to the rare effects,
 through the scaling properties  of the critical wavefunction in a similar fashion
to the Anderson transition~\cite{Pixley:2015}.
The disorder averaged inverse participation ratios (IPR's),
$P_q = \int d^d r |\psi|^{2q}$  are expected to scale with the size of the system $L$ as
$\overline{P_q} \sim L^{-\tilde{\tau}_q}$, where the $\tilde{\tau}_{q}$ describe the multifractal spectrum of the wavefunctions.
In the semi-metallic phase the only possible states
at the nodal point are the algebraically-decaying instanton-like solutions
predicted in Ref.~\cite{Nandkishore:2014} and observed numerically in Ref.~\cite{Pixley:2016}.
Since these states, if
present with finite density, are localized, we still obtain $\tilde{\tau}_q=0$ in the semi-metallic phase, at least for small $q$.
In the diffusive metal phase the system has a finite density
of extended states at zero energy  that results in  $\tilde{\tau}_q=d(q-1)$.
Exactly at the transition the exponent modifies to
$\tilde{\tau}_q=d(q-1)+ \tilde{\Delta}_q$,
where $\tilde{\Delta}_q$
also governs the scaling of the moments of the local DOS (LDOS),
$\overline{\rho^q} \sim L^{-\tilde{\Delta}_q}$.
It is related by $\tilde{\Delta}_q =  x^*_q-q x^*_1$ to the scaling dimension $x^*_q$ of the
local composite operator
representing the $q$th moment of the LDOS. Fortunately, the scaling dimension of this operator has been calculated  within the GN model
to two-loop order in Ref.~\cite{Foster:2012} and reads
$x^*_q= (d-1)q  - 2 q\Delta_0^*  - 2 \Delta_0^{*2} [3q(q-1) +q] $ with $\Delta_0^*$
obtained from~(\ref{eq-beta1}).
Note that $\tilde{\Delta}_q=\frac38 q(1-q)\varepsilon_2^2$ satisfies the convexity inequality
$\partial^2 x^*_q/\partial q^2 \le 0 $~\cite{Duplantier:1991} and the identity
$\tilde{\Delta}_q = \tilde{\Delta}_{1-q}$, i.e.
$\overline{\tilde{\rho}^q }\sim \overline{\tilde{\rho}^{1-q}}$ with $\tilde{\rho} = \rho/\overline{\rho}$.
The latter holds for the  multifractal  exponents  in  the  different Wigner-Dyson classes~\cite{Evers:2008}
and follows from a very general symmetry of the LDOS distribution
$P_{\rho} (\tilde{\rho}) = \tilde{\rho }^{-3} P_{\rho} (\tilde{\rho }^{-1})$~\cite{Mirlin:2006}. 
Then the two-point correlation function is expected to scale as
$\overline{\tilde{\rho}^p(r)  \tilde{\rho}^{q}(0)} \sim (r/a)^{-\tilde{\Delta}_p-\tilde{\Delta}_q} (r/ L)^{\tilde{\Delta}_{p+q}}$,
where $a$ is the microscopic cutoff.
Crucially, this description of the multifractal spectrum of the critical wavefunctions, at least for small $q$ is weakly sensitive
to the presence of rare events and indeed characterizes  the
underlying  avoided critical point. 

\textit{Note added.} During the final completion of this paper,  we became aware of the 
recent preprint~\cite{Syzranov:2016} where the authors independently 
came to the
 same conclusions about the multifractality at the transition. 
 
\textit{Acknowledgments.} We would like to thank V. Gurarie  for his inspiring questions and
 J.H. Pixley for drawing our attention to his recent work~\cite{Pixley:2016b}.
 We acknowledge support from the
 French Agence Nationale de la Recherche by Grant 
ANR-12-BS04-0007 (SemiTopo).

%



\newpage
\begin{widetext}
\begin{center}
{\Large \bf Supplemental Material} \\
{\large \bf On the disorder-driven quantum transition in three-dimensional relativistic metals} \\
{\large T. Louvet, D. Carpentier, and A. A. Fedorenko}
\end{center} 
\end{widetext}

\setcounter{equation}{0}
\setcounter{figure}{0}

\section{Generalized Gross-Neveu model: $ 2+\varepsilon$ expansion }
\label{sec:GN-ren}

The minimal action of the Weyl fermions in $d$ dimensions can be rewritten in Fourier
space as
\begin{eqnarray} \label{eq:action-b}
  S &=&
\sum\limits_{\alpha=1}^N \int_{k} \bar{\psi}_{\alpha}(-\mathbf{k})(  \bm{\gamma} \mathbf{k}  -
 i \omega  )   \psi_{\alpha}(\mathbf{k}) \nonumber \\
&& - \sum\limits_{\alpha,\beta=1}^N \sum\limits_{n,\vec{A}}\frac{\Delta_n}2 \int_{k_i} \ [\bar{\psi}_\alpha(\mathbf{k}_1) \gamma^{(n)}_{\vec{A}} \psi_\alpha (\mathbf{k}_2)] \nn \\
&& \ \ \ \ \ \times [\bar{\psi}_\beta(\mathbf{k}_3) \gamma^{(n)}_{\vec{A}} \psi_\beta (-\mathbf{k}_1-\mathbf{k}_2-\mathbf{k}_3)].
\end{eqnarray}
One can build up a perturbation theory in small disorder calculating  all
correlation and vertex functions  perturbatively in $\Delta_n$.
Each term can be represented as a Feynman diagram. In these diagrams the solid lines
stand for the the bare propagator
\begin{eqnarray}
&& \langle \bar{\psi}_\alpha(\mathbf{k,\omega}) \psi_{\beta}(-\mathbf{k},-\omega) \rangle_0 = \delta_{\alpha\beta}
\frac{\bm{\gamma} \mathbf{k}  + i \omega}{k^2+\omega^2},
\end{eqnarray}
and the dashed line corresponds to one of the vertex $\frac12 \Delta_n$.
Note that the dashed line transmit only momenta but not frequency.
These terms turn out to be diverging in $d=2$  which is the lower critical dimension of the transition.
Simple scaling analysis shows that weak disorder is irrelevant for $d>2$.
To make the theory finite we use the dimensional regularization and compute all integrals in $d=2+\varepsilon_2$.
At the end we put $\varepsilon_2=1$.
To render the divergences we employ the minimal subtraction scheme and collect all poles in $\varepsilon_2 $ in the Z-factors:
$Z_{\psi}$, $Z_\omega$ and $Z_n$ so that the correlation function calculated with the  renormalized action
\begin{eqnarray} \label{eq:action-r}
 && S_R =
\sum\limits_{\alpha=1}^N \int_{k} \bar{\psi}_{\alpha}(-\mathbf{k})(  Z_\psi \bm{\gamma} \mathbf{k}  -
Z_\omega i \omega  )  \psi_{\alpha}(\mathbf{k}) \nn \\
&& - \sum\limits_{\alpha,\beta=1}^N \sum\limits_{n,A} \frac{ \mu^{-\varepsilon }\Delta_n}{K_d}  \int_{k_i} \ [\bar{\psi}_\alpha \gamma^{(n)}_{A} \psi_\alpha ] \cdot [\bar{\psi}_\beta \gamma^{(n)}_{A} \psi_\beta ] \ \ \ \  \\
\end{eqnarray}
remain finite in the limit
$\varepsilon_2 \to 0$.
Here we have introduced the renormalized fermionic fields $\psi$, $\bar{\psi}$ and
the renormalized dimensionless coupling constants $\Delta_n$ on the mass scale $\mu$, which are related to the parameters by
\begin{eqnarray} \label{eq:Z-factors}
&&\mathring{\psi} = Z_{\psi}^{1/2} \psi, \ \ \ \mathring{\bar{\psi}} = Z_{\psi}^{1/2}  \bar{\psi},\\
&&\mathring{\omega} = Z_\omega Z_{\psi}^{-1}\omega, \ \ \ \
 \mathring{\Delta}_n =
 \frac{2 \mu^{-\varepsilon }}{K_d} \frac{Z_n}{Z_{\psi}^{2}} \Delta_n,
 \end{eqnarray}
where $K_d = 2\pi^{d/2}/((2\pi)^d\Gamma(d/2))$ is the area of the $d$-dimensional unite sphere divided by $(2\pi)^d$.
The renormalized and the bare vertex  and Green functions are related by
\begin{eqnarray}
&&\mathring{\Gamma}^{(\mathcal{N})}(p_i, \mathring{\omega},\mathring{\Delta}) = Z_{\psi}^{-\mathcal{N}/2}{\Gamma}^{(n)}(p_i, \omega, \Delta, \mu),
\label{eq-gam1}  \\ \label{eq-gam2}
&& \mathring{G}^{(\mathcal{N})}(p_i, \mathring{\omega},\mathring{\Delta}) = Z_{\psi}^{\mathcal{N}/2}G^{(n)}(p_i, \omega, \Delta, \mu),
\end{eqnarray}
where $\Delta$ stands for all $\Delta_n$.
Using that the bare functions $\mathring{\Gamma}^{(\mathcal{N})}$  and $\mathring{G}^{(n)}$  do not depend on the renormalization scale
$\mu$ we  take the derivative  of Eqs.~(\ref{eq-gam1}) and (\ref{eq-gam2}) with respect to $\mu$  and obtain the RG flow
equations for the renormalized Green and vertex functions:
\begin{eqnarray}
&&\left[\mu\frac{\partial}{\partial \mu} - \sum\limits_n \beta_n(\Delta)
\frac{\partial}{\partial \Delta_n} - \frac{\mathcal{N}}2 \eta_\psi(\Delta) \right. \nn \\
&& \ \ \ \ \ \ \ \ \  \left.
 - \gamma (\Delta) \omega \frac{\partial}{\partial \omega}\right] {\Gamma}^{(n)}(p_i, \omega, \Delta)=0, \label{eq-rg1-1} \\
 &&\left[\mu\frac{\partial}{\partial \mu} - \sum\limits_n \beta_n(\Delta)
\frac{\partial}{\partial \Delta} +  \frac{\mathcal{N}}2 \eta_\psi(\Delta) \right. \nn \\
&& \ \ \ \ \ \ \ \ \  \left.
 - \gamma (\Delta) \omega \frac{\partial}{\partial \omega}\right] G^{(n)}(p_i, \omega, \Delta)=0. \label{eq-rg1-2}
\end{eqnarray}
Here we have defined the scaling functions
\begin{eqnarray}
&&\beta_n(\Delta)= - \left.\mu\frac{\partial \Delta_n}{\partial \mu} \right|_{\mathring{\Delta}}, \label{eq-def-beta-fun} \\
&&\eta_\psi(\Delta)= - \sum\limits_n\beta_n(\Delta)\frac{\partial \ln Z_\psi}{\partial \Delta_n}, \\
&&\eta_\omega(\Delta)= - \sum\limits_n \beta_n(\Delta)\frac{\partial \ln Z_\omega}{\partial \Delta_n}, \\
&& \gamma(\Delta) = \eta_\omega(\Delta)- \eta_\psi(\Delta).
\end{eqnarray}
Dimensional analysis gives the following rescaling formulas
\begin{eqnarray}
\Gamma^{(\mathcal{N})}(p_i, \omega, \Delta, \mu) &=&\lambda^{-d+\mathcal{N}(d-1)/2}  \nn \\
&& \times {\Gamma}^{(\mathcal{N})}(\lambda p_i, \lambda \omega, \Delta, \lambda \mu),  \\
 G^{(\mathcal{N})}(p_i, \omega, \Delta, \mu) &=&\lambda^{d(\mathcal{N}-1)-\mathcal{N}(d-1)/2} \nn \\
&& \times G^{(\mathcal{N})}(\lambda p_i, \lambda \omega, \Delta, \lambda \mu),
\end{eqnarray}
which can be rewritten in an infinitesimal form as
\begin{eqnarray}
&&\left[\mu\frac{\partial}{\partial \mu} + \sum\limits_i p_i \frac{\partial}{\partial p_i} + \omega \frac{\partial}{\partial \omega} \right. \nn \\
&&\ \ \ \ \ \ \left.
-d+\frac{\mathcal{N}(d-1)}2 \right] {\Gamma}^{(\mathcal{N})}(p_i, \omega, \Delta)=0,  \label{eq-rg2-1}\\
&&\left[\mu\frac{\partial}{\partial \mu} + \sum\limits_i p_i \frac{\partial}{\partial p_i} + \omega \frac{\partial}{\partial \omega} \right. \nn \\
&&\ \ \ \ \ \ \left.
+d(\mathcal{N}-1)-\frac{\mathcal{N}(d-1)}2  \right] G^{(\mathcal{N})}(p_i, \omega, \Delta)=0. \nn \\  \label{eq-rg2-2}
\end{eqnarray}
Subtracting Eqs.~(\ref{eq-rg1-1}) and (\ref{eq-rg1-2}) from Eqs.~(\ref{eq-rg2-1}) and (\ref{eq-rg2-2}) we obtain
\begin{eqnarray}
&& \!\!\!\!\!\!\!\!\! \left[ \sum\limits_n \beta_n(\Delta)
\frac{\partial}{\partial \Delta_n}+\sum\limits_i p_i \frac{\partial}{\partial p_i} + (1+\gamma (\Delta)) \omega \frac{\partial}{\partial \omega}   \right. \nn \\
&&  \   \left.
  -d + \frac{\mathcal{N}}2 \left[ d-1   + \eta_\psi(\Delta) \right] \right] {\Gamma}^{(\mathcal{N})}(p_i, \omega, \Delta)=0, \nn \\ \label{eq-rg3-1} \\
&&  \!\!\!\!\!\!\!\!\! \left[\sum\limits_n \beta_n(\Delta)
\frac{\partial}{\partial \Delta_n} +\sum\limits_i p_i \frac{\partial}{\partial p_i}  + (1+\gamma (\Delta))\omega \frac{\partial}{\partial \omega}
\right. \nn \\
&&  \!\!\!\!\!\!\!\!\!  \left.  + d(\mathcal{N}-1) - \frac{\mathcal{N}}2 \left[ d-1   + \eta_\psi(\Delta) \right]
 \right] G^{(\mathcal{N})}(p_i, \omega, \Delta)=0.   \nn \\ \label{eq-rg3-2}
\end{eqnarray}
The solutions of Eqs.~(\ref{eq-rg3-1}) and (\ref{eq-rg3-2}) can be found by using the method of characteristics.
The characteristics, i.e. lines in the space of $p_i$, $\omega$, and $\Delta_n$,  parameterized by auxiliary parameter $\xi$ which below will be identified with the correlation length, can be found from the equations
\begin{eqnarray}
&&   \frac{d p_i (\xi)}{d \ln \xi} = p_i(\xi),  \\
&&   \frac{d \Delta_n (\xi)}{d \ln \xi} =  \beta_n(\Delta(\xi)),  \\
&&   \frac{d \omega (\xi)}{d \ln \xi} = [1+\gamma(\Delta(\xi))] \omega(\xi),
\end{eqnarray}
with initial conditions $\Delta_n (1)=\Delta_n$, $p_i(1)=p_i$, and $\omega (1)=\omega$.
The solution of  Eqs.~(\ref{eq-rg3-1}) and (\ref{eq-rg3-2}) then propagate along the characteristics
according to the equations
\begin{eqnarray}
&&   \frac{d M_\mathcal{N}(\xi)}{d \ln \xi} = [-d + \frac{\mathcal{N}}2 (d-1 + \eta_\psi(\Delta(\xi)))] M_\mathcal{N}(\xi),  \nn
 \\
&&   \frac{d H_\mathcal{N}(\xi)}{d \ln \xi} = \left[d(\mathcal{N}-1) - \frac{\mathcal{N}}2 ( d-1 + \eta_\psi(\Delta(\xi)) )\right] H_\mathcal{N}(\xi),  \nn \\
\end{eqnarray}
with the initial conditions $M_n(1)=H_n(1)=1$. Thus the solutions of Eqs.~(\ref{eq-rg3-1}) and (\ref{eq-rg3-2})
satisfy
\begin{eqnarray}
&& {\Gamma}^{(\mathcal{N})}(p_i, \omega, \Delta)= M_\mathcal{N}(\xi) {\Gamma}^{(\mathcal{N})}( p_i(\xi),\omega(\xi), \Delta(\xi)  ), \ \  \ \ \label{eq-rg4-1}  \\
 && G^{(\mathcal{N})}(p_i, \omega, \Delta) = H_\mathcal{N}(\xi) G^{(\mathcal{N})}( p_i(\xi),\omega(\xi), \Delta(\xi)  ). \ \ \ \ \ \label{eq-rg4-2}
\end{eqnarray}
We assume that the $\beta$-function have a fixed point (FP)
\begin{eqnarray}
\beta(\Delta^*)=0,  \label{eq:fixpoint0}
\end{eqnarray}
with a single unstable direction $\delta = \Delta - \Delta^*$, \textit{i.e.}   the stability matrix
\begin{equation}\label{smatrix}
\mathcal{M}_{nm}= \left. \frac{\partial \beta_n(\Delta)}{\partial \Delta_m}\right|_{\Delta^*},
\end{equation}
has only  one positive eigenvalue $\lambda_1^{(+)}$ associated with the  direction $\delta$.
Then the solutions~(\ref{eq-rg4-1}) and (\ref{eq-rg4-2}) in the vicinity of the
FP~(\ref{eq:fixpoint0}) can be rewritten as
\begin{eqnarray}
&&\!\!\!\!\!\!\! {\Gamma}^{(\mathcal{N})}(p_i, \omega,\delta)= \xi^{\mathcal{N} d_\psi - d } f_\mathcal{N}( p_i \xi,\omega \xi^z,  \delta \xi^{1/\nu}), \ \ \  \\
 && \!\!\!\!\!\!\!  G^{(\mathcal{N})}(p_i, \omega,\delta) = \xi^{d(\mathcal{N}-1) - n d_\psi} g_\mathcal{N}( p_i \xi ,\omega \xi^z, \delta \xi^{1/\nu}), \ \ \
\end{eqnarray}
where we defined the critical exponents $\nu$, $z$, $d_\psi$. The parameter
$\xi$ can be identified with the correlation length that gives the critical exponent
for the correlation length
\begin{eqnarray}
\xi \sim \delta^{-\nu}, \ \ \ \ \frac1{\nu}=\lambda_1^{(+)},
\end{eqnarray}
and the dynamic dynamic critical exponent
 \begin{eqnarray}
\omega \sim k^z, \ \ \ \ \ z = 1+\gamma(\Delta^*).
\end{eqnarray}
The anomalous dimension of the fields $\psi$ and $\bar{\psi}$ reads
\begin{eqnarray}
d_\psi = \frac12[d-1+\eta_\psi(\Delta^*)].
\end{eqnarray}
Note, that the exponent $\eta_\psi$ characterizes the scaling behavior of the two-point function
\begin{eqnarray}
G^{(2)}(p) = \overline{\left\langle\bar{\psi}(p) \psi(-p)\right\rangle} \sim p^{-1+\eta_\psi(\Delta^*)},
\end{eqnarray}
which can be viewed  as the momentum distribution of fermions  at the transition.

\subsection{Critical exponents to three-loop order}

To renormalize the theory we use the minimal substraction scheme
\begin{eqnarray}
&& Z_{\psi} \mathring{\Gamma}^{(2)}(p,\omega=Z_\omega Z_{\psi}^{-1}\mu,\mathring{\Delta}(\Delta)) = \mathrm{finite}, \ \ \ \ \\
&& Z^2_{\psi} \mathring{\Gamma}^{(4)}_n(p_i=0, \omega=Z_\omega Z_{\psi}^{-1} \mu,\mathring{\Delta}(\Delta)) = \mathrm{finite}, \ \ \
\end{eqnarray}
where $\mathring{\Delta}(\Delta)$ is given by Eq.~(\ref{eq:Z-factors}) and $\mathring{\Gamma}^{(4)}_n$ is the renormalized vertex $V_n$.
The three-loop corrections to the vertex $V_0$ have been many times discussed in the literature in the context of the GN model~\cite{Zinn-Justin:1986SM}. The corresponding $\beta_0$-function defined in Eq.~(\ref{eq-def-beta-fun}) reads
\begin{eqnarray}
 \beta_0 &=& - \varepsilon_2 \Delta_0  - 2 \Delta_0^2 (N-2) -  4 \Delta_0^3 (N-2) \nn \\
&& + 2 \Delta_0^4 (N-2) (N-7), \label{eq-sm-beta1}
\end{eqnarray}
where we kept the dependence on $N$.
The 24 diagrams derived from the diagram (b) shown in Fig.~\ref{fig:diagrams-GN} by permutation of the dashed line ends which were
neglected in Ref.~\cite{Roy:2016SM} generate the vertex $V_3$ \cite{Vasilev:1997SM}.
Other diagrams which one has to take into account in calculation to order of $\varepsilon_2^3$ are the diagrams (c)-(e) shown in Fig.~\ref{fig:diagrams-GN}.
These diagrams with lines corresponding to $V_0$ and $V_3$ contribute to $V_4$ and with lines corresponding to $V_0$ and $V_3$
contribute to $V_4$. Since the contributions of the diagrams (b) are of order $\Delta_0^4$ one may naively conclude that while
$\Delta_0$ is of order $\varepsilon_2$, the two over vertices $\Delta_3$ and $\Delta_4$  are of order $\varepsilon_2^3$.
Indeed, the corresponding $\beta$-functions
\begin{eqnarray}
 \beta_3 &=&-\varepsilon_2 \Delta_3 + a  \Delta_0^4 +16 \Delta_0 \Delta_4 + 8 \Delta_0\Delta_3.  \label{eq-sm-beta2} \\
 \beta_4 &=&  -\varepsilon_2 \Delta_4 -4 \Delta_0\Delta_3 -12 \Delta_0 \Delta_4,  \label{eq-sm-beta3}
\end{eqnarray}
have the fixed point
\begin{eqnarray}
\Delta_0^*&=&\frac{\varepsilon_2 }{4-N} -\frac{\varepsilon_2 ^2}{2(2-N)^2} + \frac{(1+N)\varepsilon_2 ^3}{8(2-N)^3} + O(\varepsilon_2^4), \ \ \ \label{eq-sm-fp1-n} \\
\Delta_3^*&=&\frac{a \varepsilon_2 ^3 (N-8)}{16 N (N-6) (N-2)^3}+O\left(\varepsilon_2 ^4\right), \label{eq-sm-fp2-n} \\
\Delta_4^*&=& \frac{a \varepsilon_2 ^3}{8 N (N-6) (N-2)^3}+O\left(\varepsilon_2 ^4\right) \label{eq-sm-fp3-n},
\end{eqnarray}
which has non analytic behavior in the limit $N \to 0$. Taking first the limit $N \to 0$ in the $\beta$-functions
one finds the fixed point
\begin{eqnarray}
\Delta_0^*&=&\frac{\varepsilon_2 }{4} -\frac{\varepsilon_2 ^2}{8} + \frac{\varepsilon_2 ^3}{64} + O(\varepsilon_2^4), \ \ \ \label{eq-sm-fp1} \\
\Delta_3^*&=&\frac{a \varepsilon_2 ^2}{96}-\frac{23 a \varepsilon_2 ^3}{1152}+O\left(\varepsilon_2 ^4\right), \label{eq-sm-fp2} \\
\Delta_4^*&=& -\frac{a \varepsilon_2 ^2}{384}+\frac{49 a \varepsilon_2 ^3}{9216}+O\left(\varepsilon_2 ^4\right) \label{eq-sm-fp3},
\end{eqnarray}
similar to $\sqrt{\varepsilon}$ expansion for the diluted Ising model~\cite{Shalaev:1997SM}.
The stability of the FP  can be described by the eigenvalues of the stability matrix $\frac{\partial \beta_i}{\partial \Delta_j}|_{\Delta^*},~i,j \in {0,3,4}$. Since one expects that the transition is controlled by an unstable IR FP, the stability matrix is expected
to have only one positive eigenvalue which is related to the  critical exponent $1/\nu = \lambda_{1}^{(+)}$.
The stability eigenvalues read:
\begin{eqnarray}
&&\frac1{\nu} = \lambda_{1}^{(+)} = \varepsilon_4 +\frac{\varepsilon_4 ^2}{2}+\frac{3 \varepsilon_4 ^3}{8}+O\left(\varepsilon_4
   ^4\right), \label{eq:2-eps-eigenvalues3loop} \\
&& \lambda_{2}^{(-)} = -3 \varepsilon_4 +\varepsilon_4 ^2-\frac{\varepsilon_4 ^3}{8}+O\left(\varepsilon_4
   ^4\right), \\
&& \lambda_{3}^{(-)} = -\frac{\varepsilon_4 ^2}{2}+\frac{\varepsilon_4 ^3}{16}+O\left(\varepsilon_4
   ^4\right).
\end{eqnarray}
Only the first eigenvalue~\eqref{eq:2-eps-eigenvalues3loop} associated with a single instability direction is positive.

\begin{figure}
 \includegraphics[width=6.5cm]{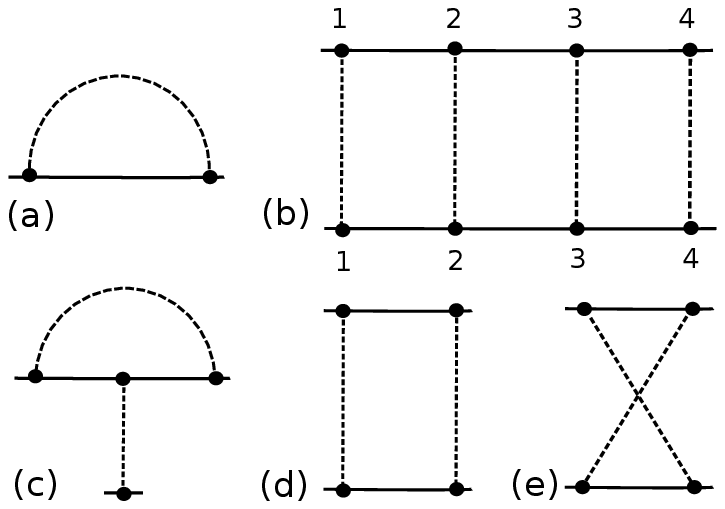}
 \caption{Diagrams entering the renormalization of the generalized GN action.
 Solid lines stands for fermionic propagators and dashed lines for disorder vertices. }
  \label{fig:diagrams-GN}
\end{figure}

The generation of vertices $\Delta_3$ and $\Delta_4$ at three-loop order might renormalize $\omega$ and thus give a correction to the
other critical exponents via diagrams of the type of diagram (a) of Fig.~\ref{fig:diagrams-GN}. The
combinatorial factor associated to this diagram is 2. The contribution will take the form ($n=3,4$):

  \begin{align}
 I_n & = \sum_{A=\{i_1,..,i_n\}}  \gamma^{(n)}_A \int_{\bf k}  \frac{\bm{\gamma} \mathbf{k}  + i \omega}{k^2+\omega^2} \gamma^{(n)}_A \nn \\
     & = \sum_{A=\{i_1,..,i_n\}} \gamma^{(n)}_A \gamma^{(n)}_A \int_{\bf k}  \frac{i \omega}{k^2+\omega^2},
     \label{eq:3loop-zdiag}
  \end{align}
besides
\begin{align}
 \gamma^{(n)}_A & = As[\gamma_{i_1}....\gamma_{i_n}] \nn \\
		& = \epsilon^{i_1..i_n} \gamma_{i_1}....\gamma_{i_n},
\end{align}
where the set of indices is set and $\epsilon^{i_1..i_n}$ is the corresponding element of the n-th Levi Civita tensor. Therefore (no contraction on A is implied here):
\begin{align}
 (\gamma^{(n)}_A)^2 & = (\epsilon^{i_1..i_n})^2 \gamma_{i_1}....\gamma_{i_n} \gamma_{i_1}....\gamma_{i_n} \nn \\
		  & = \gamma_{i_1}....\gamma_{i_n} \gamma_{i_1}....\gamma_{i_n} \nn \\
		  & = (-1)^{n-1} \gamma_{i_2}....\gamma_{i_n} (\gamma_{i_1})^2 \gamma_{i_2}....\gamma_{i_n} \nn \\
		  & = (-1)^{(n-1)!} \mathbb{I}
\end{align}
using the anticommutation relation $\{\gamma_\mu,\gamma_\nu\} = 2\delta_{\mu \nu} \mathbb{I}$ and assuming all indices $ i_1,...,i_n $ are distinct (otherwise, $\gamma^{(n)}_A$ vanishes trivially.) Note that since $n=3$ or $4$,  $(-1)^{(n-1)!} = 1$.
Performing the sum in \eqref{eq:3loop-zdiag} thus yields:
\begin{align}
 I_n & = \binom{d}{n} \int_{\bf k}  \frac{i \omega}{k^2+\omega^2} \propto (d-2) \int \frac{d^d{\bf k}}{(2\pi)^d} \frac{i \omega}{k^2+\omega^2}\nn \\
\end{align}
For $n=3,4$, $\binom{d}{n} \propto (d-2) = O (\varepsilon)$, and thus the binomial coefficient cancels the pole in the integral, making the contribution $I_n  = O(1)$ finite. At the end of the day we find that this diagram will give no contribution to the frequency renormalisation
and a fortiori to the $z$ exponent. Thus  the critical exponents  to three-loop order are given by

\begin{eqnarray}
 z &=& 1+ \frac{\varepsilon_2}{2} -\frac{\varepsilon_2^2}{8} +  \frac{3\varepsilon_2^3}{32} + O(\varepsilon_2^4) , \label{eq:exp-z}\\
 \eta &=& -\frac{\varepsilon_2^2}{8} + \frac{3\varepsilon_2^3}{16} -  \frac{25\varepsilon_2^4}{128} + O(\varepsilon_2^5). \\
 d_\psi &=& \frac12[d-1+\eta_\psi] \nn \\
& = & \frac12 + \frac{\varepsilon}{2} -\frac{\varepsilon^2}{16} +  \frac{3\varepsilon_2^3}{32} -  \frac{25\varepsilon_2^4}{256} + O(\varepsilon_2^5).
\end{eqnarray}
To estimate numerical values of the exponents in $d=3$ we use direct evaluation at $\varepsilon_2=1$ (D), Pad\'{e} approximant P$[M/L]$
and Pad\'{e}-Borel resummation PB$[M/L]$.
We find  $z=1.469$ (D), $z=1.429$ (P[2/1]) and $z=1.425$ (PB[2/1]);
$\eta=0.0625$ (D) to three loop and $\eta=-0.133$ (D) to four loop.

\subsection{Renormalization of composite operators}

We now discuss the renormalization of the composite operators
\begin{eqnarray}
\mathcal{O}_q (r) := \left(\bar{\psi}_\alpha(r){\psi}_\alpha(r)\right)^q. \label{eq:comp-oper}
\end{eqnarray}
which are related to the the deviation of the disorder distribution from the Gaussian
distribution. The bare scaling dimension of operators~(\ref{eq:comp-oper}) is $[\mathcal{O}_q ]=(d-1)q + O(\Delta)$.
To find their scaling dimension in the GN FP we introduce the $Z$-factors
\begin{eqnarray}
\mathring{\mathcal{O}}_q = \mathcal{Z}_{q} Z_\psi^{-q}{\mathcal{O}}_q. \label{eq:comp-oper-ren}
\end{eqnarray}
which has to render the divergence of the correlation functions involving operators~(\ref{eq:comp-oper}) .
To one loop order  the diagrams contributing to the $Z_q$  factor are shown in Fig.~\ref{fig:diagrams4-composite}.
We find to one-loop order
\begin{eqnarray}
\mathcal{Z}_{q} = 1 + 2[q+q(q-1)]\frac{\Delta_0}{\varepsilon_2}. \label{eq:comp-oper-ren-Z}
\end{eqnarray}
The corresponding scaling function
\begin{eqnarray}
&&\eta_q(\Delta)= - \sum\limits_n \beta_n(\Delta)\frac{\partial \ln Z_q}{\partial \Delta_n},
\end{eqnarray}
gives the scaling dimension of the composite operators~(\ref{eq:comp-oper})
\begin{eqnarray}
&&[\mathcal{O}_q ]=(d-1+\eta_\psi)q - \eta_q(\Delta^*).
\end{eqnarray}
To one loop order this yields
\begin{eqnarray}
[\mathcal{O}_q] = (1+\varepsilon_2)q - \frac12 q^2 \varepsilon_2 +O(\varepsilon_2^2),
\end{eqnarray}
that is consistent with the conformal theory results of~\cite{Ghosh:2016SM}.

\begin{figure}
 \includegraphics[width=8cm]{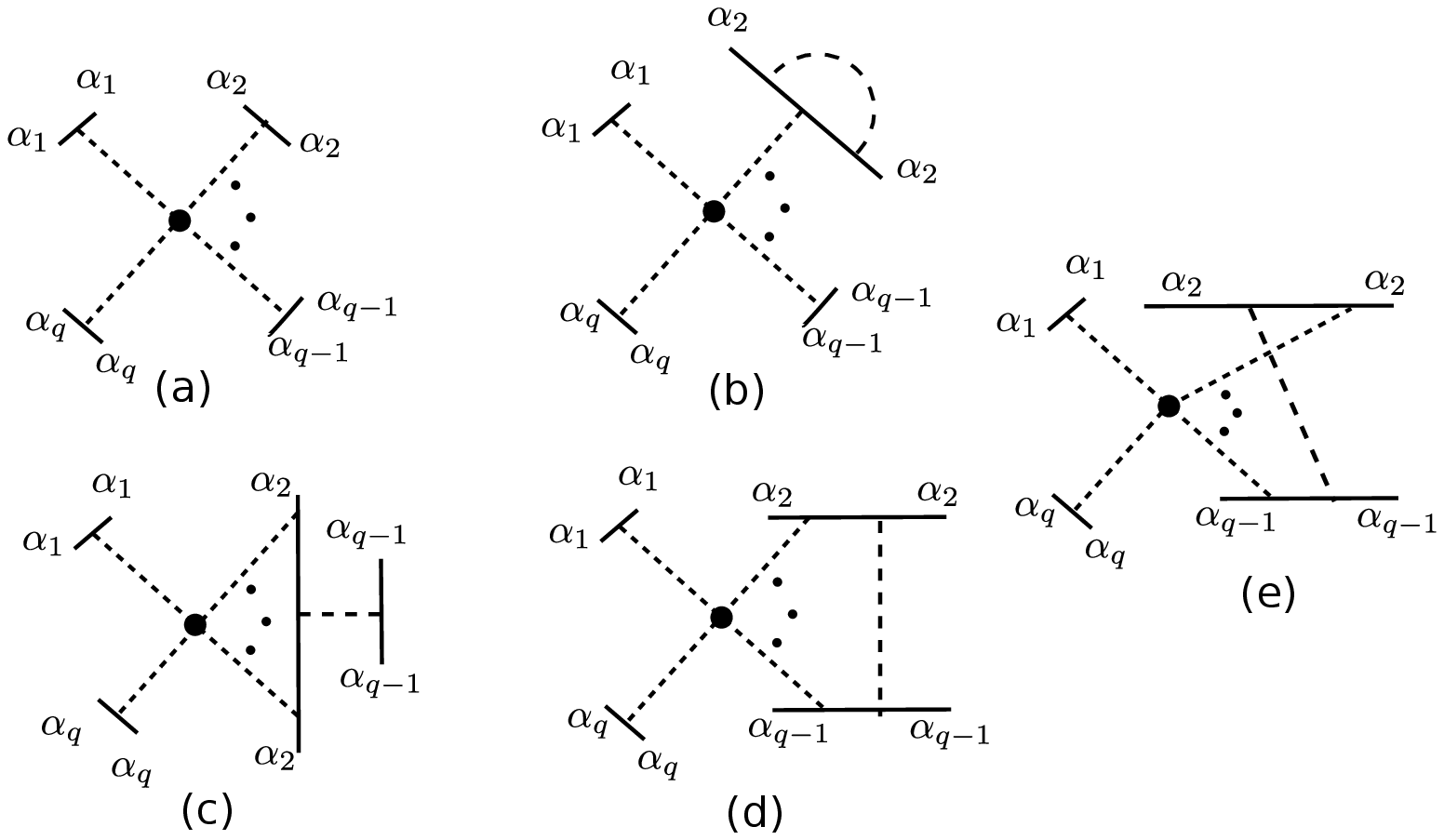}
 \caption{Diagrams renormalizing the composite operators (\ref{eq:comp-oper}). (a) is the bare vertex corresponding to a composite operator  (\ref{eq:comp-oper}), (a) - (e) are the one-loop diagrams contributing to renormalization  (\ref{eq:comp-oper-ren}): (b) $= q I_1$,
 (c)$ = q(q-1)I_1$, (d)+(e) $= 0$, where $I_1$ is the one-loop integral.   }
  \label{fig:diagrams4-composite}
\end{figure}

In order to calculate the scaling behavior of the local DOS $\rho(\omega,\delta)$ it is enough to consider renormalization
of the composite operator $\mathcal{O}_1$. The corresponding $Z$ factor is not independent
and is related to $Z_\omega$ by
\begin{eqnarray}
\mathring{\mathcal{O}}_1 = Z_\omega Z_\psi^{-1} \mathcal{O}_1.
\end{eqnarray}
We can write the flow equation for the local DOS as
\begin{eqnarray}
&&\left[ \sum\limits_n \beta_n(\Delta)
\frac{\partial}{\partial \Delta_n} + (1+\gamma (\Delta)) \omega \frac{\partial}{\partial \omega}   \right. \nn \\
&& \ \ \   \left.
   - (d-1)  +  \eta_\omega(\Delta) - \eta_\psi(\Delta)  \right]\rho(\omega,\Delta)=0. \label{eq:DOS1}
\end{eqnarray}
The solution of Eq.~(\ref{eq:DOS1}) in the vicinity of the FP~(\ref{eq:fixpoint0})  has the form
\begin{eqnarray}
 && \rho( \omega) = \xi^{z-d} \rho_0(\omega \xi^z, \delta \xi^{1/\nu}), \ \ \
\end{eqnarray}
with $z=[\mathcal{O}_1]$ given to three-loop order by (\ref{eq:exp-z}).

\section{Gross-Neveu-Yukawa Model: $4-\varepsilon$ expansion }
\label{sec:gny}
\subsection{Model}
The action for the $U(N)$ GNY model is given by
 \begin{align}
  S_{GNY} &=  \int d^d r [ - {\bar \chi}_\alpha ( {\bm \gamma} \cdot {\bm \nabla} + \sqrt{g} \phi ) \chi_\alpha  \nn \\
    &+ \frac12 (\nabla \phi)^2 + \frac12 \mu \phi^2 + \frac{\lambda}{4\text{!}} \phi^4 ].
    \label{eq:action-GNY-SM}
 \end{align}
We are interested in the $N \rightarrow 0$ limit. In Fourier space ($- i\bm{\gamma} \cdot \bm{\nabla}  \to \bm{\gamma} \cdot \mathbf{k}$) , the bare fermionic and bosonic propagators read
\begin{subequations}
 \begin{align}
 \langle \chi_\alpha ({\bf k}) \chi_\alpha (- {\bf k'}) \rangle = i \frac{ {\bm \gamma} \cdot {\bf k}}{k^2} \\
 \langle \phi ({\bf q}) \phi(- {\bf q}) \rangle = \frac{1}{q^2 + \mu} .
 \end{align}
\end{subequations}

\subsection{Renormalization}
\begin{figure}[htbp]
 \includegraphics[width=7.5cm]{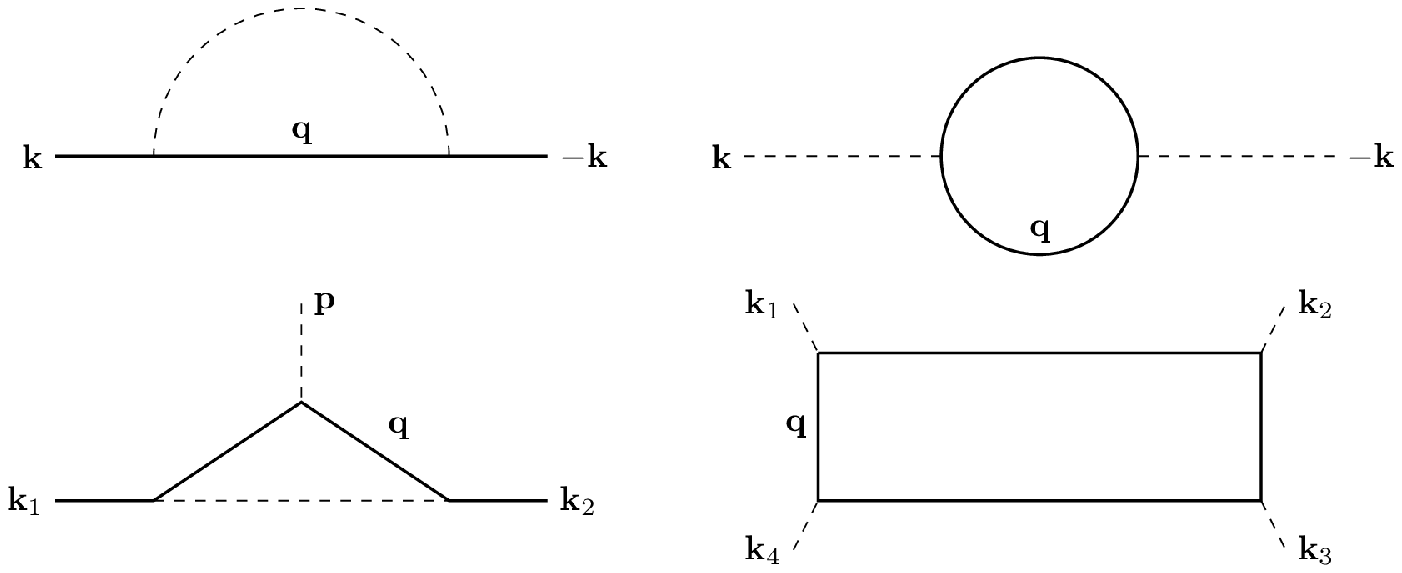}
 \caption{Diagrams entering the renormalization of the GNY action to one-loop order. Solid lines stand for the fermionic propagator
 and dashed lines for the bosonic one. Diagrams from $\phi^4$ theory not shown.}
  \label{fig:diagrams4-eps}
\end{figure}
We perform a perturbative expansion of correlation and vertex functions in the disorder parameters $g$ and $\lambda$.
Fig. \ref{fig:diagrams4-eps} shows the diverging diagrams in $d=4$ that involve fermionic-bosonic vertices. Other diverging diagrams come from the expansion in $\lambda$ and are known from the $\phi^4$ theory.
We use dimensional regularization in $d=4-\varepsilon_4$ and put $\varepsilon_4 = 1$ at the end of the day.

Following the minimal substraction scheme, we introduce the renormalization constants
$Z_\chi$, $Z_\phi$, $Z_\mu$, $Z_g$ and $Z_\lambda$. Calling $\Lambda$ the renormalization scale, the renormalized action reads:
\begin{align}
  S &=  \int d^d r [ - {\bar \chi}_\alpha (Z_\chi {\bm \gamma} \cdot {\bm \nabla}  + \Lambda^{\varepsilon_4/2} \sqrt{g Z_g} \phi ) \chi_\alpha  \nonumber \\
   &+ \frac12 Z_\phi (\nabla \phi)^2 + \frac12 [\mu_c Z_\phi  + \Lambda^2 Z_\mu ~ \delta \mu] \phi^2 + \Lambda^\varepsilon_4 Z_\lambda \frac{\lambda}{4\text{!}} \phi^4 ].
\end{align}
The renormalized fields are related to the bare ones through $\mathring{\chi} = Z_\chi^{1/2} \chi$, and $\mathring{\phi} = Z_\phi^{1/2} \phi$.
Similarly, we define the renormalized bosonic mass $\mathring{\mu} = \mu_c + \Lambda^2 Z_\mu Z_\phi^{-1} ~ \delta \mu$. The relations between bare and renormalized couplings read
$\mathring{g} = \Lambda^{\varepsilon_4} Z_g Z_\chi^{-2} Z_\phi^{-1} g $ and $\mathring{\lambda} = \Lambda^{\varepsilon_4} Z_\lambda Z_\phi^{-2} \lambda$, where we have introduced the renormalization scale $\Lambda$ to render the renormalized couplings dimensionless.

The bare and renormalized correlation and vertex functions are related as follows:
\begin{subequations}
\begin{align}
   \mathring{\Gamma}^{(n,l)} & (p_i, q_j, \mathring{\mu}-\mu_c, \mathring{g} , \mathring{\lambda})  =  \nn \\
			     & Z_\chi^{-n/2} Z_\phi^{-l/2} \Gamma^{(n,l)} (p_i, q_j, \delta\mu, g, \lambda, \Lambda ), \label{eq:sm-ren-gamma}\\
   \mathring{G}^{(n,l)} & (p_i, q_j, \mathring{\mu}-\mu_c, \mathring{g} , \mathring{\lambda})  =  \nn \\
			& Z_\chi^{n/2} Z_\phi^{l/2} G^{(n,l)} (p_i, q_j, \delta \mu, g, \lambda, \Lambda ).
\end{align}
\end{subequations}
From Eq.~(\ref{eq:sm-ren-gamma}) we derive the RG flow equation for the vertex functions:
\begin{align}
 &  \left[ \Lambda \frac{\partial}{\partial \Lambda} - \beta_{g} \frac{\partial}{\partial g} - \beta_{\lambda} \frac{\partial}{\partial \lambda} - \frac n 2 \eta_\chi - \frac l 2 \eta_\phi \right. \nn \\
   & \ \ \ \ \ \left. -  \gamma_\mu \delta\mu \frac{\partial}{\partial \delta \mu} \right] \Gamma^{(n,l)} (p_i,  q_j, \delta \mu, g, \lambda, \Lambda ) =0,
\label{eq:4-eps_rgeq-a}
\end{align}
with the scaling functions:
\begin{subequations}
 \begin{align}
 \beta_{g}(g,\lambda) & = - \left.\Lambda \frac{\partial g}{\partial \Lambda}\right|_{\mathring{g},\mathring{\lambda}}, \\
 \beta_\lambda (g,\lambda) & = - \left. \Lambda \frac{\partial \lambda}{\partial \Lambda}\right|_{\mathring{g},\mathring{\lambda}}, \\
 \eta_\chi(g,\lambda) & = - \sum_{u = \lambda,g} \beta_{u} \frac{\partial \ln Z_\chi}{\partial u}, \\
 \eta_\phi(g,\lambda) & = - \sum_{u = \lambda,g} \beta_{u} \frac{\partial \ln Z_\phi}{\partial u}, \\
 \eta_\mu(g,\lambda) & = - \sum_{u = \lambda,g} \beta_{u} \frac{ \partial \ln Z_\mu}{\partial u}, \\
 \gamma_\mu (g,\lambda) & = 2 + \eta_\mu - \eta_\phi.
 \end{align}
\end{subequations}
Besides, dimensional analysis gives
\begin{subequations}
\begin{align}
  & \Gamma^{(n,l)}  (p_i, q_j, \delta \mu, g , \lambda, \Lambda)  = X^{- d + n (d-1)/2 + l (d-2)/2 } \nn \\
		  &  \hspace{20mm} \times \Gamma^{(n,l)} (X p_i, X q_j, \delta \mu, g, \lambda, X \Lambda ),  \label{eq:sp:gamma-dim-analys}\\
  & G^{(n,l)}  (p_i, q_j, \delta \mu, g , \lambda, \Lambda)  =  X^{(n + l -1) d - n(d-1)/2 - l(d-2)/2} \nn \\
  &   \hspace{20mm} \times G^{(n,l)} (X p_i, X q_j, \delta \mu, g, \lambda, X \Lambda ).
\end{align}
\end{subequations}
We rewrite the relation (\ref{eq:sp:gamma-dim-analys}) in an infinitesimal form as
\begin{align}
 &  \left[ \Lambda \frac{\partial}{\partial \Lambda} + p_i \frac{\partial}{\partial p_i} + q_j \frac{\partial}{\partial q_j}   - d + \frac{n(d-1)}2   \right. \nn \\
   & \ \ \ \ \ \ \ \ \left.   + \frac{l(d-2)}2  \right] \Gamma^{(n,l)} (p_i, q_j, \delta \mu, g, \lambda, \Lambda ),
\label{eq:4-eps_rgeq-b}
\end{align}
Subtracting \eqref{eq:4-eps_rgeq-a} from \eqref{eq:4-eps_rgeq-b}  to get rid of the derivative with respect to $\Lambda$ we obtain
\begin{eqnarray}
 && \!\!\!\!\!\!\!\!\!\!\!\!
  \left[ \beta_{g} \frac{\partial}{\partial g} + \beta_{\lambda} \frac{\partial}{\partial \lambda} + p_i \frac{\partial}{\partial p_i} + q_j \frac{\partial}{\partial q_j} \right. \nn \\
   \nn \\
   &&  + \gamma_\mu \delta\mu \frac{\partial}{\partial \delta \mu} - d + \frac{n}2 ( \eta_\chi + d - 1)  \nn \\
   && \left.
   + \frac{l}2 ( \eta_\phi + d - 2)  \right] \Gamma^{(n,l)} (p_i, q_j, \delta \mu, g, \lambda, \Lambda ) =0.
\label{eq:4-eps_rgeq-c}
\end{eqnarray}
The solutions to Eq.~\eqref{eq:4-eps_rgeq-c} can be found using the method of characteristics. These solutions propagate along
specific lines in the space of $p_i$, $q_j$, $\delta \mu$, $g$ and $\lambda$ called the characteristics. The characteristics are parametrized by
an auxiliary parameter $L$, which can be identified with a length scale; they are determined by the following set of RG flow equations:
\begin{subequations}
 \begin{align}
  \frac{d p_i(L)}{d \ln L} & = p_i(L), \\
  \frac{d q_i(L)}{d \ln L} & = q_i(L), \\
  \frac{d \delta\mu(L)}{d \ln L} & = \gamma_\mu  \delta\mu(L), \\
  \frac{d g (L)}{d \ln L} & = \beta_{g}(g(L)) ,\\
  \frac{d \lambda (L)}{d \ln L} & = \beta_\lambda (\lambda(L)),
 \end{align}
\label{eq:RG-4-eps}
\end{subequations}
with initial conditions $p_i(1) = p_i$, $q_j(1) = q_j$, $\delta\mu(1) = \delta\mu$, $g (1) = g$, $\lambda (1) = \lambda$.
Thus the solutions of \eqref{eq:4-eps_rgeq-c} satisfy
\begin{align}
 \Gamma^{(n,l)} &  (p_i,q_i,\delta \mu,g,\lambda) =  \nn \\
 & \mathcal{M}(L) \Gamma^{(n,l)} (p_i(L), q_i(L),\delta \mu(L),g(L),\lambda(L)).
 \label{eq:4-eps_scale}
\end{align}
with
\begin{equation}
   \frac{d \ln \mathcal{M}_{n,l}(L)}{d \ln L}  = \frac{n}2 (\eta_\chi + d - 1) + \frac{l}2 ( \eta_\phi + d - 2) - d .
\end{equation}
In the vicinity of the critical point the RG flow parameter $L$ can be identified with the correlation length $\xi$ in \eqref{eq:RG-4-eps}, allowing one to calculate the critical exponents from the RG equations.

\subsection{Critical exponents}
Calculation of the one-loop diagrams shown in Fig.~\ref{fig:diagrams4-eps} in the limit $N \rightarrow 0$
gives~\cite{Zinn-Justin:1986SM}:
\begin{subequations}
\begin{align}
 \Gamma^{(2,0)} & = \langle \bar{\chi} ({\bf k}) \chi(- {\bf k}) \rangle^{-1}
		 = i {\bm \gamma}{\bm p} Z_\chi + i {\bm \gamma}{\bm p} \frac{K_d}2\frac{g}{\varepsilon_4}, \\
 \Gamma^{(2,1)} & = \langle \bar{\chi}({\bf k}_1) \chi({\bf k}_2) \phi({\bf p}) \rangle_{\mathrm{1PI}} = \sqrt{g Z_g} - g^{3/2} \frac{K_d}{\varepsilon_4}, \\
 \Gamma^{(0,2)} & = \langle \phi({\bf k }) \phi(-{\bf k}) \rangle^{-1} = Z_\phi k^2  + Z_{\mu} \delta\mu - \frac{\lambda}2 \delta\mu \frac{K_d}{\varepsilon_4}, \\
 \Gamma^{(0,4)} & = \langle \phi({\bf k }_1) \phi({\bf k}_2) \phi({\bf k}_3) \phi({\bf k}_4) \rangle_{\mathrm{1PI}} = Z_\lambda \lambda - \frac32 \lambda^2 \frac{K_d}{\varepsilon_4}.
 \end{align}
\end{subequations}
To make these functions finite, we define the renormalization constants as follows:
\begin{subequations}
 \begin{align}
  Z_\chi & = 1 - \frac12 g \frac{K_d}{\varepsilon_4}, \\
  Z_\phi & = 1, \\
  Z_{\mu} & = 1 + \frac{\lambda}{2} \frac{K_d}{\varepsilon_4}, \\
  Z_g & = 1 + 2 g \frac{K_d}{\varepsilon_4},  \\
  Z_\lambda & = 1 + \frac32 \lambda \frac{K_d}{\varepsilon_4}.
  \label{eq:GNY-counterterms}
 \end{align}
\end{subequations}
It is convenient to include $K_d/2$ in the redefinition of $g$ and $\lambda$. The $\beta$-functions read
\begin{subequations}
 \begin{align}
 \beta_g (g,\lambda) &  =\varepsilon_4 g - 6 g^2   \\
 \beta_\lambda (g,\lambda) & =\varepsilon_4 \lambda - 3 \lambda^2,
 \end{align}
\end{subequations}
The FP solution is given by
\begin{equation}
 g_* = \frac{\varepsilon_4}{6},~\lambda_* = \frac{\varepsilon_4}{3}.
\end{equation}
The total RG flow in the three parameter space is shown in Fig.\ref{fig:3DFlow-SM}.

\begin{figure}
\includegraphics[width=75mm]{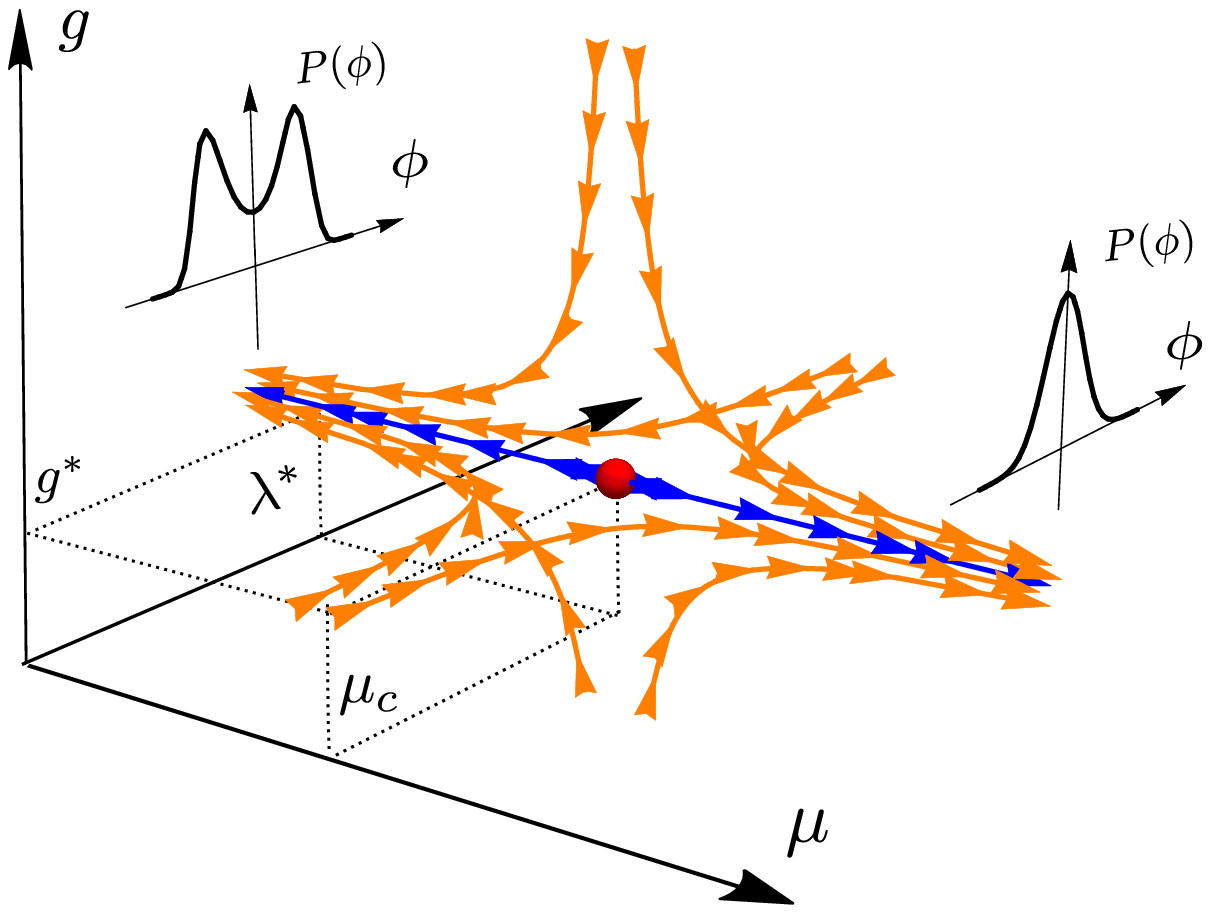}
\caption{
  \label{fig:3DFlow-SM}
 Schematic representation of the RG flow for the $U(N)$ GNY model~(\ref{eq:action-GNY-SM}) in the three-parameter space: $g$, $\lambda$ and $\mu$.
The transition is described by a fixed point of the flow
 ($g^{*}$, $\lambda^*$, $\mu_c$) which possesses only one unstable direction along $\mu$.  On one side of the transition, the  flow towards large $\mu$  corresponds to the semi-metallic phase with vanishing density of states at the band crossing and a Gaussian distribution of disorder. On the other side
 the flow towards small $\mu$ drives the system towards a diffusive metal with a finite density of states and non-Gaussian distribution of disorder. }
\end{figure}

The critical length exponent is defined by:
\begin{equation}
 \xi \sim \delta\mu^{-\nu}, \label{eq:sm-GN-nu}
\end{equation}
thus from the RG equations \eqref{eq:RG-4-eps}, identifying the parameter $L$ to the correlation length $\xi$ in the region near the critical point we get
\begin{equation}
 \nu^{-1} =\gamma_\mu =  2 - \lambda_*  = 2 - \frac{\varepsilon_4}3 + O(\varepsilon_4^2).
\end{equation}

The anomalous dimension of the fermionic field reads
\begin{align}
 d_\chi & = \frac12 ( d - 1+ \eta_\chi) = \frac12 ( d-1 + g_*) \nn \\
	& = \frac32 - \frac5{12} \varepsilon_4 + O(\varepsilon_4^2).
\end{align}
When $\delta \mu$ is negative, the scalar field acquires a finite expectation values which generates in turn a fermion mass $m_\chi  = \sqrt{g} \langle \phi \rangle$.
This fermionic mass is analogous to a frequency and thus scales with the correlation length like $m_\chi \sim \xi^{-z}$.
Besides, the correlations of the scalar field are determined by a $\phi^4$ field theory; it is known that close to the transition the order parameter of this theory scales like
\begin{align}
 \langle \phi \rangle \sim \delta\mu^\beta \sim \xi^{-\beta/\nu},
\end{align}
where  $\nu$ is given by (\ref{eq:sm-GN-nu}).
Besides, from the RG analysis we have established that the \
dimensionful coupling constant $\sqrt{g}$ flows towards the FP as $\sqrt{g} \sim \xi^{-\varepsilon_4/2} \sqrt{g^*}$. This leads to
\begin{equation}
 m_\chi \sim \sqrt{g} \langle \phi \rangle \sim \xi^{-\varepsilon_4/2-\beta/\nu},
\end{equation}
and therefore we get $z = \varepsilon_4/2 + \beta/\nu $.
Moreover, the exponent $\beta$ is related to the exponent $\nu$ through the scaling relation
\begin{equation}
  \nu d = 2\beta + (2-\eta_\phi) \nu.
\end{equation}
From \eqref{eq:GNY-counterterms} we know $Z_\phi = 1$, which gives $\eta_\phi=O(\varepsilon_4^2)$. Hence we get
\begin{equation}
 \beta = \nu \frac{2-\varepsilon_4}2,
\end{equation}
and finally we find for the critical dynamic exponent $z$ to one loop order:
\begin{equation}
 z = \frac{\varepsilon_4}2 + \frac{2-\varepsilon_4}2 = 1 + O (\varepsilon_4^2) .
\end{equation}

The two-loop order contribution can be calculated using the two loop expression of $2-\eta_\phi$~\cite{Karkkainen:1994SM}:
\begin{equation}
 2 - \eta_\phi = 2-\frac{\varepsilon_4^2}{54} +O(\varepsilon_4^3),
\end{equation}
which gives
\begin{align}
 z & = \frac{\varepsilon_4}2 + \frac{\beta}{\nu} = \frac{\varepsilon_4}2 + \frac12 (d - 2 + \eta_\phi) \nn \\
   & = 1 + \frac{\varepsilon_4^2}{108} + O(\varepsilon_4^3).
\end{align}

\subsection{Instanton solutions}

We now show the existence of localized instanton solutions to the GNY action in the limit $N \to 0$ that can give a non-zero contribution
to the zero-energy DOS in the semimetallic phase, similar to that found for the GN model in Ref.~\cite{Nandkishore:2014SM}.
 Following \cite{Falco:2009SM} we start by rewriting the average DOS at the Dirac point directly in $d=3$ in the form:
\begin{equation}
 \langle \rho (E=0) \rangle_V = \frac1{L^3} \int D[V,\chi,\Psi,\Upsilon] \exp[-S],
\end{equation}
where $\Psi(x)$ is a Lagrange multiplier field selecting solutions to the Dirac equation and $\Upsilon$ is a Lagrange multiplier enforcing normalization of $\Psi(x)$ and the action is given by
\begin{align}
 S  = & \int d^3x \left[ (\nabla \phi)^2 + \mu \phi(x)^2 + \frac{\lambda}{4!} \phi(x)^4 \right] \nn \\
   & -  \int d^3x~ \Psi^\dagger(x) (  {\bm \sigma} \cdot \nabla +  \sqrt{g} \phi ) \chi(x) \nn \\
   & + \Upsilon \left[ \int d^3x~ \chi^\dagger(x)\chi(x) - 1 \right], \label{eq:sm-action-instanton}
\end{align}
where ${\bm \sigma} = \sigma_x,\sigma_y,\sigma_z $ are the Pauli matrices.
We now look for a  saddle-point solution to the classical equations of motion. To obtain the latter
we vary the action~(\ref{eq:sm-action-instanton}) with respect to $\phi$, $\chi$, $\chi^\dagger$, $\Psi^\dagger$, and  $\Upsilon$.
\begin{center}
\begin{figure}
 \includegraphics[width=7.5cm]{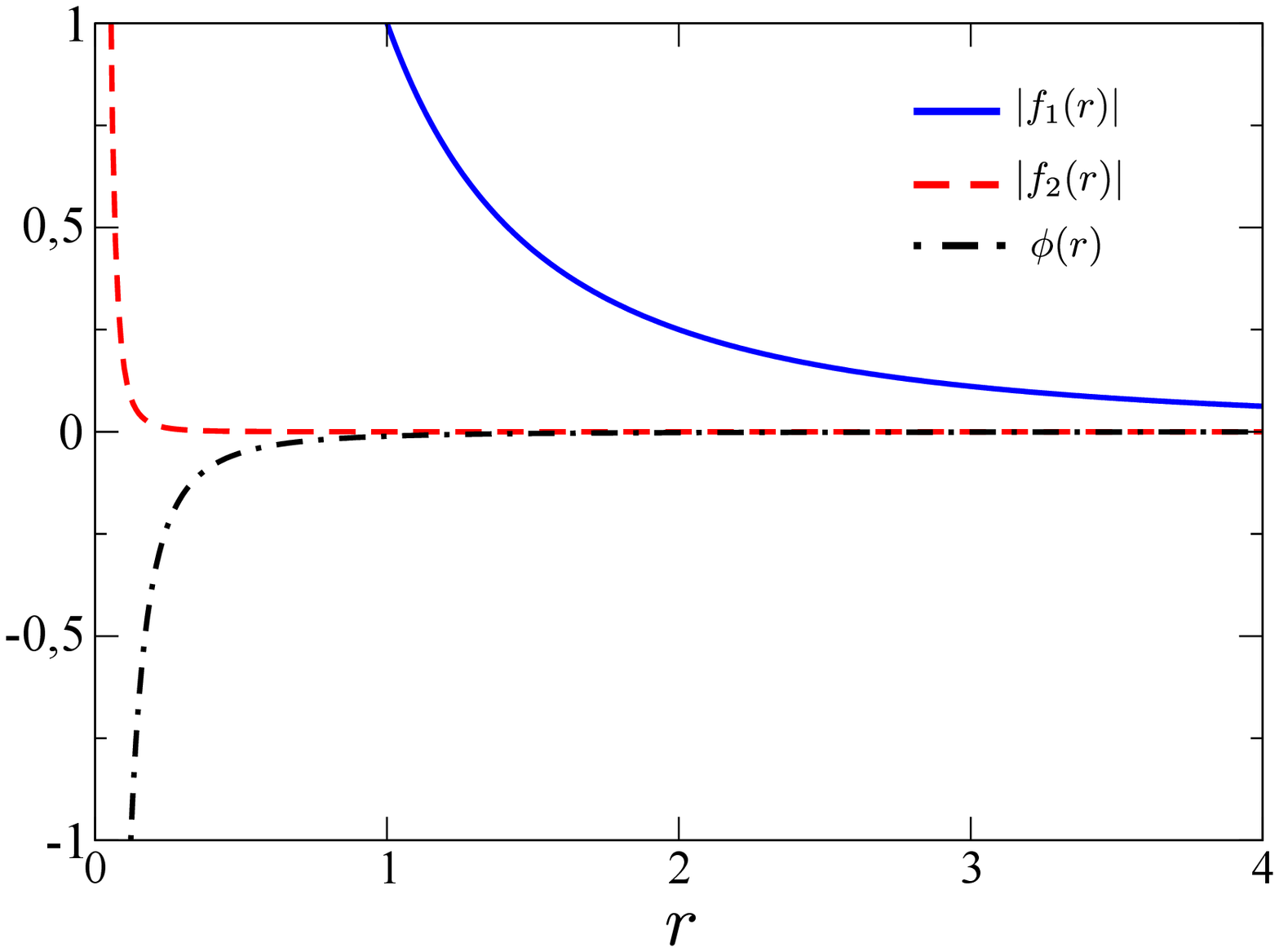}
 \caption{ The instanton wavefunction components ($f_1$: blue, continuous, $f_2$: red, dashed)
 \eqref{eq:instanton-wf} and scalar field $\phi$ (black, dot-dashed)  solution of \eqref{eq:instanton-a}-\eqref{eq:instanton} computed numerically using  the expansion \eqref{eq:instanton-expansion}-\eqref{eq:instanton-effective-pot}
 for $A=1$,  $\mu=-0.1$, $\lambda=0.001$, $g=0.001$, and  $\Psi_0=1$ . \label{fig:instantons-1}}
\end{figure}
\end{center}
This yields
\begin{equation}
 - \nabla^2 \phi(x) + \mu \phi (x) + \frac{\lambda}{3!} \phi(x)^3 = \sqrt{g} \Psi^\dagger(x) \chi(x), \label{eq-sm-ceofm1}
\end{equation}
\begin{equation}
 \Psi^\dagger(x)[{\bm \sigma} \cdot \nabla + \sqrt{g} \phi(x)] = 0,
\end{equation}
\begin{equation}
 [{\bm \sigma} \cdot \nabla + \sqrt{g} \phi(x)]\chi(x) = 0,
\end{equation}
\begin{equation}
 \int d^3x~ \chi^\dagger(x) \chi(x) = 1,
\end{equation}
\begin{equation}
 \Upsilon ~\chi(x) = 0. \label{eq-sm-ceofm10}
\end{equation}
From Eq.~(\ref{eq-sm-ceofm10}) it follows that $\Upsilon=0$ and thus we can take $\Psi^\dagger(x) = \Psi_0 \chi^\dagger (x)$ where $\Psi_0$ is a real number.
Since the disorder distribution is isotropic it is naturally to assume that the DOS is dominated by a spherically symmetric
saddle-point solution. This drastically simplifies the solution of the classical equations of motion (\ref{eq-sm-ceofm1})-(\ref{eq-sm-ceofm10}) since they can be reduced to the problem of a Dirac particle in
a self-consistent central potential.
The solution to this problem can be factorized in the radial and angular parts according to
\begin{equation}
 \chi = f_1(r) \varphi^- -f_2(r) \varphi^+,
 \label{eq:instanton-wf}
\end{equation}
where $\varphi^\pm$ are two-component spinors with total angular momentum $j$, angular momentum along $z$ $j_z$ and orbital angular momentum $l^\pm = j \mp 1/2$. We have:
\begin{align}
 {\bm \sigma} \cdot {\bm \nabla} f_i (r) \varphi^\pm = & {\bm \sigma} \cdot {\hat {\bf r}} \left( \partial_r - \frac{ {\bm \sigma} \cdot {\bf L}}{ r} \right)  f_i (r) \varphi^\pm  \nn \\
 = & \left( \partial_r + \frac{1-\kappa}r \right) f_i (r) ~  {\bm \sigma} \cdot {\hat {\bf r}}~ \varphi^\pm \nn \\
 = & \left( \partial_r + \frac{1-\kappa}r \right) f_i (r) \varphi^\mp,
\end{align}
with $\kappa = \pm (j+1/2) = \pm 1$ for the lowest level $j=1/2$. Thus we get the following system:
\begin{eqnarray}
&& \!\!\!  \partial_r f_2(r)   = f_1(r) \sqrt{g} \phi(r),  \label{eq:instanton-a}
\end{eqnarray}
\begin{eqnarray}
&& \!\!\!  (\partial_r + \frac2r) f_1(r)  =  f_2(r) \sqrt{g} \phi(r),
\end{eqnarray}
\begin{eqnarray}
&& \!\!\!   -(\partial_r^2 + \frac2r \partial_r - \mu) \phi  + \frac{\lambda}{3!} \phi^3 = \sqrt{g} \Psi_0 \left[ |f_1|^2 + |f_2|^2 \right]. \ \ \ \
  \label{eq:instanton}
\end{eqnarray}
The large $r$ expansion of the Eqs.~(\ref{eq:instanton-a})-(\ref{eq:instanton}) gives the following asymptotic behavior
\begin{eqnarray}
 f_1(r) & = &\frac{A}{r^2} + \frac{A^5 g^2}{30 \mu^2} \frac1{r^8} + O(\frac{A^9 g^4}{\mu^4 r^{14}}), \label{eq:instanton-expansion}  \\
 f_2 (r) & = &- \frac{A^3 g}{5\mu} \frac1{r^5} - 7 \frac{A^7 g^3}{550 \mu^3} \frac1{r^{11}} + O(\frac{A^{11} g^5}{\mu^5 r^{17}}),
 \\
  \phi (r) & = &\frac{A^2 \sqrt{g}}{\mu} \frac1{r^4} + O(\frac{A^6 g^{5/2}}{\mu^3 r^{10}}),
 \label{eq:instanton-effective-pot}
\end{eqnarray}
which depends on a single free parameter $A$.
A typical solution to Eqs.~(\ref{eq:instanton-a})-(\ref{eq:instanton}) obtained numerically using the asymptotic
behavior~(\ref{eq:instanton-expansion})-(\ref{eq:instanton-effective-pot}) is shown in  Fig.\ref{fig:instantons-1}.
The wave function and disorder distribution both exhibit a singular behavior at $r=0$ and thus require a regularization~\cite{Nandkishore:2014SM}.
Moreover, to obtain the full instanton contribution to the DOS (using either GN or GNY models) one has to expand around
the instanton solution and calculate the corresponding Gaussian integral which gives a prefactor to the exponential behavior. It is known
that in the case of quadratic dispersion this prefactor can be expressed in the form of a ratio of two functional determinants which
diverges in $d>1$. Thus, in this case the instanton solution requires renormalization~\cite{Falco:2015SM}.
The regularization and renormalization of the instanton solution in the case of a Dirac particles in disordered potential is an
interesting open question which is left for the future.


%


\begin{thebibliography}{50}%
\makeatletter
\providecommand \@ifxundefined [1]{%
 \@ifx{#1\undefined}
}%
\providecommand \@ifnum [1]{%
 \ifnum #1\expandafter \@firstoftwo
 \else \expandafter \@secondoftwo
 \fi
}%
\providecommand \@ifx [1]{%
 \ifx #1\expandafter \@firstoftwo
 \else \expandafter \@secondoftwo
 \fi
}%
\providecommand \natexlab [1]{#1}%
\providecommand \enquote  [1]{``#1''}%
\providecommand \bibnamefont  [1]{#1}%
\providecommand \bibfnamefont [1]{#1}%
\providecommand \citenamefont [1]{#1}%
\providecommand \href@noop [0]{\@secondoftwo}%
\providecommand \href [0]{\begingroup \@sanitize@url \@href}%
\providecommand \@href[1]{\@@startlink{#1}\@@href}%
\providecommand \@@href[1]{\endgroup#1\@@endlink}%
\providecommand \@sanitize@url [0]{\catcode `\\12\catcode `\$12\catcode
  `\&12\catcode `\#12\catcode `\^12\catcode `\_12\catcode `\%12\relax}%
\providecommand \@@startlink[1]{}%
\providecommand \@@endlink[0]{}%
\providecommand \url  [0]{\begingroup\@sanitize@url \@url }%
\providecommand \@url [1]{\endgroup\@href {#1}{\urlprefix }}%
\providecommand \urlprefix  [0]{URL }%
\providecommand \Eprint [0]{\href }%
\providecommand \doibase [0]{http://dx.doi.org/}%
\providecommand \selectlanguage [0]{\@gobble}%
\providecommand \bibinfo  [0]{\@secondoftwo}%
\providecommand \bibfield  [0]{\@secondoftwo}%
\providecommand \translation [1]{[#1]}%
\providecommand \BibitemOpen [0]{}%
\providecommand \bibitemStop [0]{}%
\providecommand \bibitemNoStop [0]{.\EOS\space}%
\providecommand \EOS [0]{\spacefactor3000\relax}%
\providecommand \BibitemShut  [1]{\csname bibitem#1\endcsname}%
\let\auto@bib@innerbib\@empty
\bibitem [{\citenamefont {Liu}\ \emph {et~al.}(2014)\citenamefont {Liu},
  \citenamefont {Zhou}, \citenamefont {Zhang}, \citenamefont {Wang},
  \citenamefont {Weng}, \citenamefont {Prabhakaran}, \citenamefont {Mo},
  \citenamefont {Shen}, \citenamefont {Fang}, \citenamefont {Dai},
  \citenamefont {Hussain},\ and\ \citenamefont {Chen}}]{Liu:2014}%
  \BibitemOpen
  \bibfield  {author} {\bibinfo {author} {\bibfnamefont {Z.~K.}\ \bibnamefont
  {Liu}}, \bibinfo {author} {\bibfnamefont {B.}~\bibnamefont {Zhou}}, \bibinfo
  {author} {\bibfnamefont {Y.}~\bibnamefont {Zhang}}, \bibinfo {author}
  {\bibfnamefont {Z.~J.}\ \bibnamefont {Wang}}, \bibinfo {author}
  {\bibfnamefont {H.~M.}\ \bibnamefont {Weng}}, \bibinfo {author}
  {\bibfnamefont {D.}~\bibnamefont {Prabhakaran}}, \bibinfo {author}
  {\bibfnamefont {S.~K.}\ \bibnamefont {Mo}}, \bibinfo {author} {\bibfnamefont
  {Z.~X.}\ \bibnamefont {Shen}}, \bibinfo {author} {\bibfnamefont
  {Z.}~\bibnamefont {Fang}}, \bibinfo {author} {\bibfnamefont {X.}~\bibnamefont
  {Dai}}, \bibinfo {author} {\bibfnamefont {Z.}~\bibnamefont {Hussain}}, \ and\
  \bibinfo {author} {\bibfnamefont {Y.~L.}\ \bibnamefont {Chen}},\ }\href@noop
  {} {\bibfield  {journal} {\bibinfo  {journal} {Science}\ }\textbf {\bibinfo
  {volume} {343}},\ \bibinfo {pages} {864} (\bibinfo {year}
  {2014})}\BibitemShut {NoStop}%
\bibitem [{\citenamefont {Neupane}\ \emph {et~al.}(2014)\citenamefont
  {Neupane}, \citenamefont {Xu}, \citenamefont {Sankar}, \citenamefont
  {Alidoust}, \citenamefont {Bian}, \citenamefont {Liu}, \citenamefont
  {Belopolski}, \citenamefont {Chang}, \citenamefont {Jeng}, \citenamefont
  {Lin}, \citenamefont {Bansil}, \citenamefont {Chou},\ and\ \citenamefont
  {Hasan}}]{Neupane:2014}%
  \BibitemOpen
  \bibfield  {author} {\bibinfo {author} {\bibfnamefont {M.}~\bibnamefont
  {Neupane}}, \bibinfo {author} {\bibfnamefont {S.-Y.}\ \bibnamefont {Xu}},
  \bibinfo {author} {\bibfnamefont {R.}~\bibnamefont {Sankar}}, \bibinfo
  {author} {\bibfnamefont {N.}~\bibnamefont {Alidoust}}, \bibinfo {author}
  {\bibfnamefont {G.}~\bibnamefont {Bian}}, \bibinfo {author} {\bibfnamefont
  {C.}~\bibnamefont {Liu}}, \bibinfo {author} {\bibfnamefont {I.}~\bibnamefont
  {Belopolski}}, \bibinfo {author} {\bibfnamefont {T.-R.}\ \bibnamefont
  {Chang}}, \bibinfo {author} {\bibfnamefont {H.-T.}\ \bibnamefont {Jeng}},
  \bibinfo {author} {\bibfnamefont {H.}~\bibnamefont {Lin}}, \bibinfo {author}
  {\bibfnamefont {A.}~\bibnamefont {Bansil}}, \bibinfo {author} {\bibfnamefont
  {F.}~\bibnamefont {Chou}}, \ and\ \bibinfo {author} {\bibfnamefont {M.~Z.}\
  \bibnamefont {Hasan}},\ }\href@noop {} {\bibfield  {journal} {\bibinfo
  {journal} {Nature Communications}\ }\textbf {\bibinfo {volume} {5}},\
  \bibinfo {pages} {3786} (\bibinfo {year} {2014})}\BibitemShut {NoStop}%
\bibitem [{\citenamefont {Borisenko}\ \emph {et~al.}(2014)\citenamefont
  {Borisenko}, \citenamefont {Gibson}, \citenamefont {Evtushinsky},
  \citenamefont {Zabolotnyy}, \citenamefont {B{\"u}chner},\ and\ \citenamefont
  {Cava}}]{Borisenko:2014}%
  \BibitemOpen
  \bibfield  {author} {\bibinfo {author} {\bibfnamefont {S.}~\bibnamefont
  {Borisenko}}, \bibinfo {author} {\bibfnamefont {Q.}~\bibnamefont {Gibson}},
  \bibinfo {author} {\bibfnamefont {D.}~\bibnamefont {Evtushinsky}}, \bibinfo
  {author} {\bibfnamefont {V.}~\bibnamefont {Zabolotnyy}}, \bibinfo {author}
  {\bibfnamefont {B.}~\bibnamefont {B{\"u}chner}}, \ and\ \bibinfo {author}
  {\bibfnamefont {R.~J.}\ \bibnamefont {Cava}},\ }\href@noop {} {\bibfield
  {journal} {\bibinfo  {journal} {Phys. Rev. Lett.}\ }\textbf {\bibinfo
  {volume} {113}},\ \bibinfo {pages} {027603} (\bibinfo {year}
  {2014})}\BibitemShut {NoStop}%
\bibitem [{\citenamefont {Xu}\ \emph {et~al.}(2015{\natexlab{a}})\citenamefont
  {Xu}, \citenamefont {Belopolski}, \citenamefont {Alidoust}, \citenamefont
  {Neupane}, \citenamefont {Bian}, \citenamefont {Zhang}, \citenamefont
  {Sankar}, \citenamefont {Chang}, \citenamefont {Yuan}, \citenamefont {Lee},
  \citenamefont {Huang}, \citenamefont {Zheng}, \citenamefont {Ma},
  \citenamefont {Sanchez}, \citenamefont {Wang}, \citenamefont {Bansil},
  \citenamefont {Chou}, \citenamefont {Shibayev}, \citenamefont {Lin},
  \citenamefont {Jia},\ and\ \citenamefont {Hasan}}]{Xu:2015a}%
  \BibitemOpen
  \bibfield  {author} {\bibinfo {author} {\bibfnamefont {S.-Y.}\ \bibnamefont
  {Xu}}, \bibinfo {author} {\bibfnamefont {I.}~\bibnamefont {Belopolski}},
  \bibinfo {author} {\bibfnamefont {N.}~\bibnamefont {Alidoust}}, \bibinfo
  {author} {\bibfnamefont {M.}~\bibnamefont {Neupane}}, \bibinfo {author}
  {\bibfnamefont {G.}~\bibnamefont {Bian}}, \bibinfo {author} {\bibfnamefont
  {C.}~\bibnamefont {Zhang}}, \bibinfo {author} {\bibfnamefont
  {R.}~\bibnamefont {Sankar}}, \bibinfo {author} {\bibfnamefont
  {G.}~\bibnamefont {Chang}}, \bibinfo {author} {\bibfnamefont
  {Z.}~\bibnamefont {Yuan}}, \bibinfo {author} {\bibfnamefont {C.-C.}\
  \bibnamefont {Lee}}, \bibinfo {author} {\bibfnamefont {S.-M.}\ \bibnamefont
  {Huang}}, \bibinfo {author} {\bibfnamefont {H.}~\bibnamefont {Zheng}},
  \bibinfo {author} {\bibfnamefont {J.}~\bibnamefont {Ma}}, \bibinfo {author}
  {\bibfnamefont {D.~S.}\ \bibnamefont {Sanchez}}, \bibinfo {author}
  {\bibfnamefont {B.}~\bibnamefont {Wang}}, \bibinfo {author} {\bibfnamefont
  {A.}~\bibnamefont {Bansil}}, \bibinfo {author} {\bibfnamefont
  {F.}~\bibnamefont {Chou}}, \bibinfo {author} {\bibfnamefont {P.~P.}\
  \bibnamefont {Shibayev}}, \bibinfo {author} {\bibfnamefont {H.}~\bibnamefont
  {Lin}}, \bibinfo {author} {\bibfnamefont {S.}~\bibnamefont {Jia}}, \ and\
  \bibinfo {author} {\bibfnamefont {M.~Z.}\ \bibnamefont {Hasan}},\ }\href@noop
  {} {\bibfield  {journal} {\bibinfo  {journal} {Science}\ }\textbf {\bibinfo
  {volume} {349}},\ \bibinfo {pages} {613} (\bibinfo {year}
  {2015}{\natexlab{a}})}\BibitemShut {NoStop}%
\bibitem [{\citenamefont {Xu}\ \emph {et~al.}(2015{\natexlab{b}})\citenamefont
  {Xu}, \citenamefont {Alidoust}, \citenamefont {Belopolski}, \citenamefont
  {Yuan}, \citenamefont {Bian}, \citenamefont {Chang}, \citenamefont {Zheng},
  \citenamefont {Strocov}, \citenamefont {Sanchez}, \citenamefont {Chang},
  \citenamefont {Zhang}, \citenamefont {Mou}, \citenamefont {Wu}, \citenamefont
  {Huang}, \citenamefont {Lee}, \citenamefont {Huang}, \citenamefont {Wang},
  \citenamefont {Bansil}, \citenamefont {Jeng}, \citenamefont {Neupert},
  \citenamefont {Kaminski}, \citenamefont {Lin}, \citenamefont {Jia},\ and\
  \citenamefont {Zahid~Hasan}}]{Xu:2015b}%
  \BibitemOpen
  \bibfield  {author} {\bibinfo {author} {\bibfnamefont {S.-Y.}\ \bibnamefont
  {Xu}}, \bibinfo {author} {\bibfnamefont {N.}~\bibnamefont {Alidoust}},
  \bibinfo {author} {\bibfnamefont {I.}~\bibnamefont {Belopolski}}, \bibinfo
  {author} {\bibfnamefont {Z.}~\bibnamefont {Yuan}}, \bibinfo {author}
  {\bibfnamefont {G.}~\bibnamefont {Bian}}, \bibinfo {author} {\bibfnamefont
  {T.-R.}\ \bibnamefont {Chang}}, \bibinfo {author} {\bibfnamefont
  {H.}~\bibnamefont {Zheng}}, \bibinfo {author} {\bibfnamefont {V.~N.}\
  \bibnamefont {Strocov}}, \bibinfo {author} {\bibfnamefont {D.~S.}\
  \bibnamefont {Sanchez}}, \bibinfo {author} {\bibfnamefont {G.}~\bibnamefont
  {Chang}}, \bibinfo {author} {\bibfnamefont {C.}~\bibnamefont {Zhang}},
  \bibinfo {author} {\bibfnamefont {D.}~\bibnamefont {Mou}}, \bibinfo {author}
  {\bibfnamefont {Y.}~\bibnamefont {Wu}}, \bibinfo {author} {\bibfnamefont
  {L.}~\bibnamefont {Huang}}, \bibinfo {author} {\bibfnamefont {C.-C.}\
  \bibnamefont {Lee}}, \bibinfo {author} {\bibfnamefont {S.-M.}\ \bibnamefont
  {Huang}}, \bibinfo {author} {\bibfnamefont {B.}~\bibnamefont {Wang}},
  \bibinfo {author} {\bibfnamefont {A.}~\bibnamefont {Bansil}}, \bibinfo
  {author} {\bibfnamefont {H.-T.}\ \bibnamefont {Jeng}}, \bibinfo {author}
  {\bibfnamefont {T.}~\bibnamefont {Neupert}}, \bibinfo {author} {\bibfnamefont
  {A.}~\bibnamefont {Kaminski}}, \bibinfo {author} {\bibfnamefont
  {H.}~\bibnamefont {Lin}}, \bibinfo {author} {\bibfnamefont {S.}~\bibnamefont
  {Jia}}, \ and\ \bibinfo {author} {\bibfnamefont {M.}~\bibnamefont
  {Zahid~Hasan}},\ }\href@noop {} {\bibfield  {journal} {\bibinfo  {journal}
  {Nat Phys}\ }\textbf {\bibinfo {volume} {11}},\ \bibinfo {pages} {748}
  (\bibinfo {year} {2015}{\natexlab{b}})}\BibitemShut {NoStop}%
\bibitem [{\citenamefont {Goswami}\ and\ \citenamefont
  {Chakravarty}(2011)}]{Goswami:2011}%
  \BibitemOpen
  \bibfield  {author} {\bibinfo {author} {\bibfnamefont {P.}~\bibnamefont
  {Goswami}}\ and\ \bibinfo {author} {\bibfnamefont {S.}~\bibnamefont
  {Chakravarty}},\ }\href@noop {} {\bibfield  {journal} {\bibinfo  {journal}
  {Phys. Rev. Lett.}\ }\textbf {\bibinfo {volume} {107}},\ \bibinfo {pages}
  {196803} (\bibinfo {year} {2011})}\BibitemShut {NoStop}%
\bibitem [{\citenamefont {Hosur}\ \emph {et~al.}(2012)\citenamefont {Hosur},
  \citenamefont {Parameswaran},\ and\ \citenamefont {Vishwanath}}]{Hosur:2012}%
  \BibitemOpen
  \bibfield  {author} {\bibinfo {author} {\bibfnamefont {P.}~\bibnamefont
  {Hosur}}, \bibinfo {author} {\bibfnamefont {S.~A.}\ \bibnamefont
  {Parameswaran}}, \ and\ \bibinfo {author} {\bibfnamefont {A.}~\bibnamefont
  {Vishwanath}},\ }\href {\doibase 10.1103/PhysRevLett.108.046602} {\bibfield
  {journal} {\bibinfo  {journal} {Phys. Rev. Lett.}\ }\textbf {\bibinfo
  {volume} {108}},\ \bibinfo {pages} {046602} (\bibinfo {year}
  {2012})}\BibitemShut {NoStop}%
\bibitem [{\citenamefont {Ominato}\ and\ \citenamefont
  {Koshino}(2014)}]{Ominato:2014}%
  \BibitemOpen
  \bibfield  {author} {\bibinfo {author} {\bibfnamefont {Y.}~\bibnamefont
  {Ominato}}\ and\ \bibinfo {author} {\bibfnamefont {M.}~\bibnamefont
  {Koshino}},\ }\href {\doibase 10.1103/PhysRevB.89.054202} {\bibfield
  {journal} {\bibinfo  {journal} {Phys. Rev. B}\ }\textbf {\bibinfo {volume}
  {89}},\ \bibinfo {pages} {054202} (\bibinfo {year} {2014})}\BibitemShut
  {NoStop}%
\bibitem [{\citenamefont {Nandkishore}\ \emph {et~al.}(2014)\citenamefont
  {Nandkishore}, \citenamefont {Huse},\ and\ \citenamefont
  {Sondhi}}]{Nandkishore:2014}%
  \BibitemOpen
  \bibfield  {author} {\bibinfo {author} {\bibfnamefont {R.}~\bibnamefont
  {Nandkishore}}, \bibinfo {author} {\bibfnamefont {D.~A.}\ \bibnamefont
  {Huse}}, \ and\ \bibinfo {author} {\bibfnamefont {S.~L.}\ \bibnamefont
  {Sondhi}},\ }\href@noop {} {\bibfield  {journal} {\bibinfo  {journal} {Phys.
  Rev. B}\ }\textbf {\bibinfo {volume} {89}},\ \bibinfo {pages} {245110}
  (\bibinfo {year} {2014})}\BibitemShut {NoStop}%
\bibitem [{\citenamefont {Pixley}\ \emph
  {et~al.}(2016{\natexlab{a}})\citenamefont {Pixley}, \citenamefont {Huse},\
  and\ \citenamefont {Das~Sarma}}]{Pixley:2016}%
  \BibitemOpen
  \bibfield  {author} {\bibinfo {author} {\bibfnamefont {J.~H.}\ \bibnamefont
  {Pixley}}, \bibinfo {author} {\bibfnamefont {D.~A.}\ \bibnamefont {Huse}}, \
  and\ \bibinfo {author} {\bibfnamefont {S.}~\bibnamefont {Das~Sarma}},\ }\href
  {\doibase 10.1103/PhysRevX.6.021042} {\bibfield  {journal} {\bibinfo
  {journal} {Phys. Rev. X}\ }\textbf {\bibinfo {volume} {6}},\ \bibinfo {pages}
  {021042} (\bibinfo {year} {2016}{\natexlab{a}})}\BibitemShut {NoStop}%
\bibitem [{\citenamefont {Roy}\ \emph {et~al.}(2016)\citenamefont {Roy},
  \citenamefont {Juricic},\ and\ \citenamefont {Sarma}}]{Roy:2016b}%
  \BibitemOpen
  \bibfield  {author} {\bibinfo {author} {\bibfnamefont {B.}~\bibnamefont
  {Roy}}, \bibinfo {author} {\bibfnamefont {V.}~\bibnamefont {Juricic}}, \ and\
  \bibinfo {author} {\bibfnamefont {S.~D.}\ \bibnamefont {Sarma}},\ }\href@noop
  {} {} (\bibinfo {year} {2016}),\ \bibinfo {note}
  {arXiv:1603.00017}\BibitemShut {NoStop}%
\bibitem [{\citenamefont {Sbierski}\ \emph {et~al.}(2014)\citenamefont
  {Sbierski}, \citenamefont {Pohl}, \citenamefont {Bergholtz},\ and\
  \citenamefont {Brouwer}}]{Sbierski:2014}%
  \BibitemOpen
  \bibfield  {author} {\bibinfo {author} {\bibfnamefont {B.}~\bibnamefont
  {Sbierski}}, \bibinfo {author} {\bibfnamefont {G.}~\bibnamefont {Pohl}},
  \bibinfo {author} {\bibfnamefont {E.~J.}\ \bibnamefont {Bergholtz}}, \ and\
  \bibinfo {author} {\bibfnamefont {P.~W.}\ \bibnamefont {Brouwer}},\ }\href
  {\doibase 10.1103/PhysRevLett.113.026602} {\bibfield  {journal} {\bibinfo
  {journal} {Phys. Rev. Lett.}\ }\textbf {\bibinfo {volume} {113}},\ \bibinfo
  {pages} {026602} (\bibinfo {year} {2014})}\BibitemShut {NoStop}%
\bibitem [{\citenamefont {Fradkin}(1986)}]{Fradkin:1986}%
  \BibitemOpen
  \bibfield  {author} {\bibinfo {author} {\bibfnamefont {E.}~\bibnamefont
  {Fradkin}},\ }\href@noop {} {\bibfield  {journal} {\bibinfo  {journal} {Phys.
  Rev. B}\ }\textbf {\bibinfo {volume} {33}},\ \bibinfo {pages} {3263}
  (\bibinfo {year} {1986})}\BibitemShut {NoStop}%
\bibitem [{\citenamefont {Kobayashi}\ \emph {et~al.}(2014)\citenamefont
  {Kobayashi}, \citenamefont {Ohtsuki}, \citenamefont {Imura},\ and\
  \citenamefont {Herbut}}]{Kobayashi:2014}%
  \BibitemOpen
  \bibfield  {author} {\bibinfo {author} {\bibfnamefont {K.}~\bibnamefont
  {Kobayashi}}, \bibinfo {author} {\bibfnamefont {T.}~\bibnamefont {Ohtsuki}},
  \bibinfo {author} {\bibfnamefont {K.-I.}\ \bibnamefont {Imura}}, \ and\
  \bibinfo {author} {\bibfnamefont {I.~F.}\ \bibnamefont {Herbut}},\
  }\href@noop {} {\bibfield  {journal} {\bibinfo  {journal} {Phys. Rev. Lett.}\
  }\textbf {\bibinfo {volume} {112}},\ \bibinfo {pages} {016402} (\bibinfo
  {year} {2014})}\BibitemShut {NoStop}%
\bibitem [{\citenamefont {B.Sbierski}\ \emph {et~al.}(2015)\citenamefont
  {B.Sbierski}, \citenamefont {Bergholtz},\ and\ \citenamefont
  {Brouwer}}]{Sbierski:2015}%
  \BibitemOpen
  \bibfield  {author} {\bibinfo {author} {\bibnamefont {B.Sbierski}}, \bibinfo
  {author} {\bibfnamefont {E.~J.}\ \bibnamefont {Bergholtz}}, \ and\ \bibinfo
  {author} {\bibfnamefont {P.~W.}\ \bibnamefont {Brouwer}},\ }\href@noop {}
  {\bibfield  {journal} {\bibinfo  {journal} {Phys. Rev. B}\ }\textbf {\bibinfo
  {volume} {92}},\ \bibinfo {pages} {115145} (\bibinfo {year}
  {2015})}\BibitemShut {NoStop}%
\bibitem [{\citenamefont {Chen}\ \emph {et~al.}(2015)\citenamefont {Chen},
  \citenamefont {Song}, \citenamefont {Jiang}, \citenamefont {feng Sun},
  \citenamefont {Wang},\ and\ \citenamefont {Xie}}]{Chen:2015}%
  \BibitemOpen
  \bibfield  {author} {\bibinfo {author} {\bibfnamefont {C.-Z.}\ \bibnamefont
  {Chen}}, \bibinfo {author} {\bibfnamefont {J.}~\bibnamefont {Song}}, \bibinfo
  {author} {\bibfnamefont {H.}~\bibnamefont {Jiang}}, \bibinfo {author}
  {\bibfnamefont {Q.}~\bibnamefont {feng Sun}}, \bibinfo {author}
  {\bibfnamefont {Z.}~\bibnamefont {Wang}}, \ and\ \bibinfo {author}
  {\bibfnamefont {X.}~\bibnamefont {Xie}},\ }\href@noop {} {\bibfield
  {journal} {\bibinfo  {journal} {Phys. Rev. Lett.}\ }\textbf {\bibinfo
  {volume} {115}},\ \bibinfo {pages} {246603} (\bibinfo {year}
  {2015})}\BibitemShut {NoStop}%
\bibitem [{\citenamefont {Altland}\ and\ \citenamefont
  {Bagrets}(2015)}]{Altland:2015}%
  \BibitemOpen
  \bibfield  {author} {\bibinfo {author} {\bibfnamefont {A.}~\bibnamefont
  {Altland}}\ and\ \bibinfo {author} {\bibfnamefont {D.}~\bibnamefont
  {Bagrets}},\ }\href {\doibase 10.1103/PhysRevLett.114.257201} {\bibfield
  {journal} {\bibinfo  {journal} {Phys. Rev. Lett.}\ }\textbf {\bibinfo
  {volume} {114}},\ \bibinfo {pages} {257201} (\bibinfo {year}
  {2015})}\BibitemShut {NoStop}%
\bibitem [{\citenamefont {Altland}\ and\ \citenamefont
  {Bagrets}(2016)}]{Altland:2016}%
  \BibitemOpen
  \bibfield  {author} {\bibinfo {author} {\bibfnamefont {A.}~\bibnamefont
  {Altland}}\ and\ \bibinfo {author} {\bibfnamefont {D.}~\bibnamefont
  {Bagrets}},\ }\href {\doibase 10.1103/PhysRevB.93.075113} {\bibfield
  {journal} {\bibinfo  {journal} {Phys. Rev. B}\ }\textbf {\bibinfo {volume}
  {93}},\ \bibinfo {pages} {075113} (\bibinfo {year} {2016})}\BibitemShut
  {NoStop}%
\bibitem [{\citenamefont {Syzranov}\ \emph
  {et~al.}(2015{\natexlab{a}})\citenamefont {Syzranov}, \citenamefont
  {Radzihovsky},\ and\ \citenamefont {Gurarie}}]{Syzranov:2015}%
  \BibitemOpen
  \bibfield  {author} {\bibinfo {author} {\bibfnamefont {S.~V.}\ \bibnamefont
  {Syzranov}}, \bibinfo {author} {\bibfnamefont {L.}~\bibnamefont
  {Radzihovsky}}, \ and\ \bibinfo {author} {\bibfnamefont {V.}~\bibnamefont
  {Gurarie}},\ }\href@noop {} {\bibfield  {journal} {\bibinfo  {journal} {Phys.
  Rev. Lett.}\ }\textbf {\bibinfo {volume} {114}},\ \bibinfo {pages} {166601}
  (\bibinfo {year} {2015}{\natexlab{a}})}\BibitemShut {NoStop}%
\bibitem [{\citenamefont {Syzranov}\ \emph
  {et~al.}(2015{\natexlab{b}})\citenamefont {Syzranov}, \citenamefont
  {Gurarie},\ and\ \citenamefont {Radzihovsky}}]{Syzranov:2015c}%
  \BibitemOpen
  \bibfield  {author} {\bibinfo {author} {\bibfnamefont {S.~V.}\ \bibnamefont
  {Syzranov}}, \bibinfo {author} {\bibfnamefont {V.}~\bibnamefont {Gurarie}}, \
  and\ \bibinfo {author} {\bibfnamefont {L.}~\bibnamefont {Radzihovsky}},\
  }\href {\doibase 10.1103/PhysRevB.91.035133} {\bibfield  {journal} {\bibinfo
  {journal} {Phys. Rev. B}\ }\textbf {\bibinfo {volume} {91}},\ \bibinfo
  {pages} {035133} (\bibinfo {year} {2015}{\natexlab{b}})}\BibitemShut
  {NoStop}%
\bibitem [{\citenamefont {G\"arttner}\ \emph {et~al.}(2015)\citenamefont
  {G\"arttner}, \citenamefont {Syzranov}, \citenamefont {Rey}, \citenamefont
  {Gurarie},\ and\ \citenamefont {Radzihovsky}}]{Garttner:2015}%
  \BibitemOpen
  \bibfield  {author} {\bibinfo {author} {\bibfnamefont {M.}~\bibnamefont
  {G\"arttner}}, \bibinfo {author} {\bibfnamefont {S.~V.}\ \bibnamefont
  {Syzranov}}, \bibinfo {author} {\bibfnamefont {A.~M.}\ \bibnamefont {Rey}},
  \bibinfo {author} {\bibfnamefont {V.}~\bibnamefont {Gurarie}}, \ and\
  \bibinfo {author} {\bibfnamefont {L.}~\bibnamefont {Radzihovsky}},\ }\href
  {\doibase 10.1103/PhysRevB.92.041406} {\bibfield  {journal} {\bibinfo
  {journal} {Phys. Rev. B}\ }\textbf {\bibinfo {volume} {92}},\ \bibinfo
  {pages} {041406} (\bibinfo {year} {2015})}\BibitemShut {NoStop}%
\bibitem [{\citenamefont {Syzranov}\ \emph
  {et~al.}(2016{\natexlab{a}})\citenamefont {Syzranov}, \citenamefont
  {Ostrovsky}, \citenamefont {Gurarie},\ and\ \citenamefont
  {Radzihovsky}}]{Syzranov:2015b}%
  \BibitemOpen
  \bibfield  {author} {\bibinfo {author} {\bibfnamefont {S.~V.}\ \bibnamefont
  {Syzranov}}, \bibinfo {author} {\bibfnamefont {P.~M.}\ \bibnamefont
  {Ostrovsky}}, \bibinfo {author} {\bibfnamefont {V.}~\bibnamefont {Gurarie}},
  \ and\ \bibinfo {author} {\bibfnamefont {L.}~\bibnamefont {Radzihovsky}},\
  }\href {\doibase 10.1103/PhysRevB.93.155113} {\bibfield  {journal} {\bibinfo
  {journal} {Phys. Rev. B}\ }\textbf {\bibinfo {volume} {93}},\ \bibinfo
  {pages} {155113} (\bibinfo {year} {2016}{\natexlab{a}})}\BibitemShut
  {NoStop}%
\bibitem [{\citenamefont {Roy}\ and\ \citenamefont {Sarma}(2014)}]{Roy:2014}%
  \BibitemOpen
  \bibfield  {author} {\bibinfo {author} {\bibfnamefont {B.}~\bibnamefont
  {Roy}}\ and\ \bibinfo {author} {\bibfnamefont {S.~D.}\ \bibnamefont
  {Sarma}},\ }\href@noop {} {\bibfield  {journal} {\bibinfo  {journal} {Phys.
  Rev. B}\ }\textbf {\bibinfo {volume} {90}},\ \bibinfo {pages} {241112(R)}
  (\bibinfo {year} {2014})}\BibitemShut {NoStop}%
\bibitem [{\citenamefont {Roy}\ and\ \citenamefont {Sarma}(2016)}]{Roy:2016}%
  \BibitemOpen
  \bibfield  {author} {\bibinfo {author} {\bibfnamefont {B.}~\bibnamefont
  {Roy}}\ and\ \bibinfo {author} {\bibfnamefont {S.~D.}\ \bibnamefont
  {Sarma}},\ }\href@noop {} {\bibfield  {journal} {\bibinfo  {journal} {Phys.
  Rev. B}\ }\textbf {\bibinfo {volume} {93}},\ \bibinfo {pages} {119911(E)}
  (\bibinfo {year} {2016})}\BibitemShut {NoStop}%
\bibitem [{\citenamefont {Zinn-Justin}(1986)}]{Zinn-Justin:1986}%
  \BibitemOpen
  \bibfield  {author} {\bibinfo {author} {\bibfnamefont {J.}~\bibnamefont
  {Zinn-Justin}},\ }\href@noop {} {\emph {\bibinfo {title} {Quantum field
  theory and critical phenomena}}}\ (\bibinfo  {publisher} {Clarendon Press},\
  \bibinfo {address} {Oxford},\ \bibinfo {year} {1986})\BibitemShut {NoStop}%
\bibitem [{\citenamefont {Hasenfratz}\ and\ \citenamefont
  {et~al}(1991)}]{Hasenfratz:1991}%
  \BibitemOpen
  \bibfield  {author} {\bibinfo {author} {\bibfnamefont {A.}~\bibnamefont
  {Hasenfratz}}\ and\ \bibinfo {author} {\bibnamefont {et~al}},\ }\href@noop {}
  {\bibfield  {journal} {\bibinfo  {journal} {Nucl. Phys. B}\ }\textbf
  {\bibinfo {volume} {365}},\ \bibinfo {pages} {79} (\bibinfo {year}
  {1991})}\BibitemShut {NoStop}%
\bibitem [{\citenamefont {Ludwig}\ \emph {et~al.}(1994)\citenamefont {Ludwig},
  \citenamefont {Fisher}, \citenamefont {Shankar},\ and\ \citenamefont
  {Grinstein}}]{Ludwig:1994}%
  \BibitemOpen
  \bibfield  {author} {\bibinfo {author} {\bibfnamefont {A.~W.~W.}\
  \bibnamefont {Ludwig}}, \bibinfo {author} {\bibfnamefont {M.~P.~A.}\
  \bibnamefont {Fisher}}, \bibinfo {author} {\bibfnamefont {R.}~\bibnamefont
  {Shankar}}, \ and\ \bibinfo {author} {\bibfnamefont {G.}~\bibnamefont
  {Grinstein}},\ }\href@noop {} {\bibfield  {journal} {\bibinfo  {journal}
  {Phys. Rev. B}\ }\textbf {\bibinfo {volume} {50}},\ \bibinfo {pages} {7526}
  (\bibinfo {year} {1994})}\BibitemShut {NoStop}%
\bibitem [{\citenamefont {Fedorenko}\ \emph {et~al.}(2012)\citenamefont
  {Fedorenko}, \citenamefont {Carpentier},\ and\ \citenamefont
  {Orignac}}]{Fedorenko:2012}%
  \BibitemOpen
  \bibfield  {author} {\bibinfo {author} {\bibfnamefont {A.~A.}\ \bibnamefont
  {Fedorenko}}, \bibinfo {author} {\bibfnamefont {D.}~\bibnamefont
  {Carpentier}}, \ and\ \bibinfo {author} {\bibfnamefont {E.}~\bibnamefont
  {Orignac}},\ }\href@noop {} {\bibfield  {journal} {\bibinfo  {journal} {Phys.
  Rev. B}\ }\textbf {\bibinfo {volume} {85}},\ \bibinfo {pages} {125437}
  (\bibinfo {year} {2012})}\BibitemShut {NoStop}%
\bibitem [{\citenamefont {Schuessler}\ \emph {et~al.}(2009)\citenamefont
  {Schuessler}, \citenamefont {Ostrovsky}, \citenamefont {Gornyi},\ and\
  \citenamefont {Mirlin}}]{Schuessler:2009}%
  \BibitemOpen
  \bibfield  {author} {\bibinfo {author} {\bibfnamefont {A.}~\bibnamefont
  {Schuessler}}, \bibinfo {author} {\bibfnamefont {P.~M.}\ \bibnamefont
  {Ostrovsky}}, \bibinfo {author} {\bibfnamefont {I.~V.}\ \bibnamefont
  {Gornyi}}, \ and\ \bibinfo {author} {\bibfnamefont {A.~D.}\ \bibnamefont
  {Mirlin}},\ }\href@noop {} {\bibfield  {journal} {\bibinfo  {journal} {Phys.
  Rev. B}\ }\textbf {\bibinfo {volume} {79}},\ \bibinfo {pages} {075405}
  (\bibinfo {year} {2009})}\BibitemShut {NoStop}%
\bibitem [{\citenamefont {Vasilev}\ and\ \citenamefont
  {Vyazovsky}(1997)}]{Vasilev:1997}%
  \BibitemOpen
  \bibfield  {author} {\bibinfo {author} {\bibfnamefont {A.~N.}\ \bibnamefont
  {Vasilev}}\ and\ \bibinfo {author} {\bibfnamefont {M.~I.}\ \bibnamefont
  {Vyazovsky}},\ }\href@noop {} {\bibfield  {journal} {\bibinfo  {journal}
  {Theor. Math. Phys.}\ }\textbf {\bibinfo {volume} {113}},\ \bibinfo {pages}
  {1277} (\bibinfo {year} {1997})}\BibitemShut {NoStop}%
\bibitem [{Note1()}]{Note1}%
  \BibitemOpen
  \bibinfo {note} {Other resummation methods give $\nu = 0.333$ by Pad\'{e}
  [2/1] and $\nu = 0.375$ by Pad\'{e} [1/2]. Note that the Pad\'{e}-Borel[2/1]
  has a pole, but the principal value integral gives $\nu =0.57$.}\BibitemShut
  {Stop}%
\bibitem [{\citenamefont {Gracey}(2008)}]{Gracey:2008}%
  \BibitemOpen
  \bibfield  {author} {\bibinfo {author} {\bibfnamefont {J.~A.}\ \bibnamefont
  {Gracey}},\ }\href@noop {} {\bibfield  {journal} {\bibinfo  {journal}
  {Nucl.Phys. B}\ }\textbf {\bibinfo {volume} {802}},\ \bibinfo {pages} {330}
  (\bibinfo {year} {2008})}\BibitemShut {NoStop}%
\bibitem [{Note2()}]{Note2}%
  \BibitemOpen
  \bibinfo {note} {Similarly to going from \protect \textup {\hbox
  {\mathsurround \z@ \protect \normalfont (\ignorespaces \ref
  {eq:action1}\unskip \@@italiccorr )}} to \protect \textup {\hbox
  {\mathsurround \z@ \protect \normalfont (\ignorespaces \ref
  {eq:action-naiveGN}\unskip \@@italiccorr )}}, the coupling between fermions
  of different frequencies is irrelevant for constant $\omega $ properties in
  the limit $N\to 0$.}\BibitemShut {Stop}%
\bibitem [{\citenamefont {Karkkainen}\ \emph {et~al.}(1994)\citenamefont
  {Karkkainen}, \citenamefont {Lacaze}, \citenamefont {Lacock},\ and\
  \citenamefont {B.Petersson}}]{Karkkainen:1994}%
  \BibitemOpen
  \bibfield  {author} {\bibinfo {author} {\bibfnamefont {L.}~\bibnamefont
  {Karkkainen}}, \bibinfo {author} {\bibfnamefont {R.}~\bibnamefont {Lacaze}},
  \bibinfo {author} {\bibfnamefont {P.}~\bibnamefont {Lacock}}, \ and\ \bibinfo
  {author} {\bibnamefont {B.Petersson}},\ }\href@noop {} {\bibfield  {journal}
  {\bibinfo  {journal} {Nucl. Phys. B}\ }\textbf {\bibinfo {volume} {415}},\
  \bibinfo {pages} {781} (\bibinfo {year} {1994})}\BibitemShut {NoStop}%
\bibitem [{\citenamefont {Louvet}\ \emph {et~al.}(2016)\citenamefont {Louvet},
  \citenamefont {Fedorenko},\ and\ \citenamefont {Carpentier}}]{Supplementary}%
  \BibitemOpen
  \bibfield  {author} {\bibinfo {author} {\bibfnamefont {T.}~\bibnamefont
  {Louvet}}, \bibinfo {author} {\bibfnamefont {D.}\ \bibnamefont
  {Carpentier}}, \ and\ \bibinfo {author} {\bibfnamefont {A.~A.}~\bibnamefont
  {Fedorenko}},\ }\href@noop {} {\enquote {\bibinfo {title} {Supplemental
  material}}\ } (\bibinfo {year} {see page 6})\BibitemShut {NoStop}%
\bibitem [{Note3()}]{Note3}%
  \BibitemOpen
  \bibinfo {note} {Resummation methods give $\nu =0.67$ (Pad\'{e} [1/1]), $\nu
  = 0.699$ (Pad\'{e}-Borel [1/1]).}\BibitemShut {Stop}%
\bibitem [{\citenamefont {Carpentier}\ and\ \citenamefont
  {Le~Doussal}(2001)}]{Carpentier:2000}%
  \BibitemOpen
  \bibfield  {author} {\bibinfo {author} {\bibfnamefont {D.}~\bibnamefont
  {Carpentier}}\ and\ \bibinfo {author} {\bibfnamefont {P.}~\bibnamefont
  {Le~Doussal}},\ }\href {\doibase 10.1103/PhysRevE.63.026110} {\bibfield
  {journal} {\bibinfo  {journal} {Phys. Rev. E}\ }\textbf {\bibinfo {volume}
  {63}},\ \bibinfo {pages} {026110} (\bibinfo {year} {2001})}\BibitemShut
  {NoStop}%
\bibitem [{\citenamefont {Chayes}\ \emph {et~al.}(1986)\citenamefont {Chayes},
  \citenamefont {Chayes}, \citenamefont {Fisher},\ and\ \citenamefont
  {Spencer}}]{Chayes:1986}%
  \BibitemOpen
  \bibfield  {author} {\bibinfo {author} {\bibfnamefont {J.~T.}\ \bibnamefont
  {Chayes}}, \bibinfo {author} {\bibfnamefont {L.}~\bibnamefont {Chayes}},
  \bibinfo {author} {\bibfnamefont {D.~S.}\ \bibnamefont {Fisher}}, \ and\
  \bibinfo {author} {\bibfnamefont {T.}~\bibnamefont {Spencer}},\ }\href@noop
  {} {\bibfield  {journal} {\bibinfo  {journal} {Phys. Rev. Lett.}\ }\textbf
  {\bibinfo {volume} {57}},\ \bibinfo {pages} {2999} (\bibinfo {year}
  {1986})}\BibitemShut {NoStop}%
\bibitem [{\citenamefont {P\'{a}zm\'{a}ndi}\ \emph {et~al.}(1997)\citenamefont
  {P\'{a}zm\'{a}ndi}, \citenamefont {Scalettar},\ and\ \citenamefont
  {Zim\'{a}nyi}}]{Pazmandi:1997}%
  \BibitemOpen
  \bibfield  {author} {\bibinfo {author} {\bibfnamefont {F.}~\bibnamefont
  {P\'{a}zm\'{a}ndi}}, \bibinfo {author} {\bibfnamefont {R.~T.}\ \bibnamefont
  {Scalettar}}, \ and\ \bibinfo {author} {\bibfnamefont {G.~T.}\ \bibnamefont
  {Zim\'{a}nyi}},\ }\href@noop {} {\bibfield  {journal} {\bibinfo  {journal}
  {Phys. Rev. Lett.}\ }\textbf {\bibinfo {volume} {79}},\ \bibinfo {pages}
  {5130} (\bibinfo {year} {1997})}\BibitemShut {NoStop}%
\bibitem [{\citenamefont {Pixley}\ \emph {et~al.}(2015)\citenamefont {Pixley},
  \citenamefont {Goswami},\ and\ \citenamefont {Das~Sarma}}]{Pixley:2015}%
  \BibitemOpen
  \bibfield  {author} {\bibinfo {author} {\bibfnamefont {J.~H.}\ \bibnamefont
  {Pixley}}, \bibinfo {author} {\bibfnamefont {P.}~\bibnamefont {Goswami}}, \
  and\ \bibinfo {author} {\bibfnamefont {S.}~\bibnamefont {Das~Sarma}},\ }\href
  {\doibase 10.1103/PhysRevLett.115.076601} {\bibfield  {journal} {\bibinfo
  {journal} {Phys. Rev. Lett.}\ }\textbf {\bibinfo {volume} {115}},\ \bibinfo
  {pages} {076601} (\bibinfo {year} {2015})}\BibitemShut {NoStop}%
\bibitem [{\citenamefont {Foster}(2012)}]{Foster:2012}%
  \BibitemOpen
  \bibfield  {author} {\bibinfo {author} {\bibfnamefont {M.~S.}\ \bibnamefont
  {Foster}},\ }\href@noop {} {\bibfield  {journal} {\bibinfo  {journal} {Phys.
  Rev. B}\ }\textbf {\bibinfo {volume} {85}},\ \bibinfo {pages} {085122}
  (\bibinfo {year} {2012})}\BibitemShut {NoStop}%
\bibitem [{\citenamefont {Duplantier}\ and\ \citenamefont
  {Ludwig}(1991)}]{Duplantier:1991}%
  \BibitemOpen
  \bibfield  {author} {\bibinfo {author} {\bibfnamefont {B.}~\bibnamefont
  {Duplantier}}\ and\ \bibinfo {author} {\bibfnamefont {A.~W.~W.}\ \bibnamefont
  {Ludwig}},\ }\href@noop {} {\bibfield  {journal} {\bibinfo  {journal} {Phys.
  Rev. Lett.}\ }\textbf {\bibinfo {volume} {66}},\ \bibinfo {pages} {247}
  (\bibinfo {year} {1991})}\BibitemShut {NoStop}%
\bibitem [{\citenamefont {Evers}\ and\ \citenamefont
  {Mirlin}(2008)}]{Evers:2008}%
  \BibitemOpen
  \bibfield  {author} {\bibinfo {author} {\bibfnamefont {F.}~\bibnamefont
  {Evers}}\ and\ \bibinfo {author} {\bibfnamefont {A.~D.}\ \bibnamefont
  {Mirlin}},\ }\href@noop {} {\bibfield  {journal} {\bibinfo  {journal} {Rev.
  Mod. Phys.}\ }\textbf {\bibinfo {volume} {80}},\ \bibinfo {pages} {1355}
  (\bibinfo {year} {2008})}\BibitemShut {NoStop}%
\bibitem [{\citenamefont {Mirlin}\ \emph {et~al.}(2006)\citenamefont {Mirlin},
  \citenamefont {Fyodorov}, \citenamefont {Mildenberger},\ and\ \citenamefont
  {Evers}}]{Mirlin:2006}%
  \BibitemOpen
  \bibfield  {author} {\bibinfo {author} {\bibfnamefont {A.~D.}\ \bibnamefont
  {Mirlin}}, \bibinfo {author} {\bibfnamefont {Y.~V.}\ \bibnamefont
  {Fyodorov}}, \bibinfo {author} {\bibfnamefont {A.}~\bibnamefont
  {Mildenberger}}, \ and\ \bibinfo {author} {\bibfnamefont {F.}~\bibnamefont
  {Evers}},\ }\href@noop {} {\bibfield  {journal} {\bibinfo  {journal} {Phys.
  Rev. Lett.}\ }\textbf {\bibinfo {volume} {97}},\ \bibinfo {pages} {046803}
  (\bibinfo {year} {2006})}\BibitemShut {NoStop}%
\bibitem [{\citenamefont {Syzranov}\ \emph
  {et~al.}(2016{\natexlab{b}})\citenamefont {Syzranov}, \citenamefont
  {Gurarie},\ and\ \citenamefont {Radzihovsky}}]{Syzranov:2016}%
  \BibitemOpen
  \bibfield  {author} {\bibinfo {author} {\bibfnamefont {S.~V.}\ \bibnamefont
  {Syzranov}}, \bibinfo {author} {\bibfnamefont {V.}~\bibnamefont {Gurarie}}, \
  and\ \bibinfo {author} {\bibfnamefont {L.}~\bibnamefont {Radzihovsky}},\
  }\href@noop {} {} (\bibinfo {year} {2016}{\natexlab{b}}),\ \bibinfo {note}
  {arXiv:1604.07947}\BibitemShut {NoStop}%
\bibitem [{\citenamefont {Pixley}\ \emph
  {et~al.}(2016{\natexlab{b}})\citenamefont {Pixley}, \citenamefont {Goswami},\
  and\ \citenamefont {Das~Sarma}}]{Pixley:2016b}%
  \BibitemOpen
  \bibfield  {author} {\bibinfo {author} {\bibfnamefont {J.~H.}\ \bibnamefont
  {Pixley}}, \bibinfo {author} {\bibfnamefont {P.}~\bibnamefont {Goswami}}, \
  and\ \bibinfo {author} {\bibfnamefont {S.}~\bibnamefont {Das~Sarma}},\ }\href
  {\doibase 10.1103/PhysRevB.93.085103} {\bibfield  {journal} {\bibinfo
  {journal} {Phys. Rev. B}\ }\textbf {\bibinfo {volume} {93}},\ \bibinfo
  {pages} {085103} (\bibinfo {year} {2016}{\natexlab{b}})}\BibitemShut
  {NoStop}%
\end{thebibliography}

\begin{thebibliography}{9}%
\makeatletter
\providecommand \@ifxundefined [1]{%
 \@ifx{#1\undefined}
}%
\providecommand \@ifnum [1]{%
 \ifnum #1\expandafter \@firstoftwo
 \else \expandafter \@secondoftwo
 \fi
}%
\providecommand \@ifx [1]{%
 \ifx #1\expandafter \@firstoftwo
 \else \expandafter \@secondoftwo
 \fi
}%
\providecommand \natexlab [1]{#1}%
\providecommand \enquote  [1]{``#1''}%
\providecommand \bibnamefont  [1]{#1}%
\providecommand \bibfnamefont [1]{#1}%
\providecommand \citenamefont [1]{#1}%
\providecommand \href@noop [0]{\@secondoftwo}%
\providecommand \href [0]{\begingroup \@sanitize@url \@href}%
\providecommand \@href[1]{\@@startlink{#1}\@@href}%
\providecommand \@@href[1]{\endgroup#1\@@endlink}%
\providecommand \@sanitize@url [0]{\catcode `\\12\catcode `\$12\catcode
  `\&12\catcode `\#12\catcode `\^12\catcode `\_12\catcode `\%12\relax}%
\providecommand \@@startlink[1]{}%
\providecommand \@@endlink[0]{}%
\providecommand \url  [0]{\begingroup\@sanitize@url \@url }%
\providecommand \@url [1]{\endgroup\@href {#1}{\urlprefix }}%
\providecommand \urlprefix  [0]{URL }%
\providecommand \Eprint [0]{\href }%
\providecommand \doibase [0]{http://dx.doi.org/}%
\providecommand \selectlanguage [0]{\@gobble}%
\providecommand \bibinfo  [0]{\@secondoftwo}%
\providecommand \bibfield  [0]{\@secondoftwo}%
\providecommand \translation [1]{[#1]}%
\providecommand \BibitemOpen [0]{}%
\providecommand \bibitemStop [0]{}%
\providecommand \bibitemNoStop [0]{.\EOS\space}%
\providecommand \EOS [0]{\spacefactor3000\relax}%
\providecommand \BibitemShut  [1]{\csname bibitem#1\endcsname}%
\let\auto@bib@innerbib\@empty
\bibitem [{\citenamefont {Zinn-Justin}(1986)}]{Zinn-Justin:1986SM}%
  \BibitemOpen
  \bibfield  {author} {\bibinfo {author} {\bibfnamefont {J.}~\bibnamefont
  {Zinn-Justin}},\ }\href@noop {} {\emph {\bibinfo {title} {Quantum field
  theory and critical phenomena}}}\ (\bibinfo  {publisher} {Clarendon Press},\
  \bibinfo {address} {Oxford},\ \bibinfo {year} {1986})\BibitemShut {NoStop}%
\bibitem [{\citenamefont {Roy}\ and\ \citenamefont {Sarma}(2016)}]{Roy:2016SM}%
  \BibitemOpen
  \bibfield  {author} {\bibinfo {author} {\bibfnamefont {B.}~\bibnamefont
  {Roy}}\ and\ \bibinfo {author} {\bibfnamefont {S.~D.}\ \bibnamefont
  {Sarma}},\ }\href@noop {} {\bibfield  {journal} {\bibinfo  {journal} {Phys.
  Rev. B}\ }\textbf {\bibinfo {volume} {93}},\ \bibinfo {pages} {119911}
  (\bibinfo {year} {2016})}\BibitemShut {NoStop}%
\bibitem [{\citenamefont {Vasilev}\ and\ \citenamefont
  {Vyazovsky}(1997)}]{Vasilev:1997SM}%
  \BibitemOpen
  \bibfield  {author} {\bibinfo {author} {\bibfnamefont {A.~N.}\ \bibnamefont
  {Vasilev}}\ and\ \bibinfo {author} {\bibfnamefont {M.~I.}\ \bibnamefont
  {Vyazovsky}},\ }\href@noop {} {\bibfield  {journal} {\bibinfo  {journal}
  {Theor. Math. Phys.}\ }\textbf {\bibinfo {volume} {113}},\ \bibinfo {pages}
  {1277} (\bibinfo {year} {1997})}\BibitemShut {NoStop}%
\bibitem [{\citenamefont {Shalaev}\ \emph {et~al.}(1997)\citenamefont
  {Shalaev}, \citenamefont {Antonenko},\ and\ \citenamefont
  {Sokolov}}]{Shalaev:1997SM}%
  \BibitemOpen
  \bibfield  {author} {\bibinfo {author} {\bibfnamefont {B.~N.}\ \bibnamefont
  {Shalaev}}, \bibinfo {author} {\bibfnamefont {S.~A.}\ \bibnamefont
  {Antonenko}}, \ and\ \bibinfo {author} {\bibfnamefont {A.}~\bibnamefont
  {Sokolov}},\ }\href@noop {} {\bibfield  {journal} {\bibinfo  {journal} {Phys.
  Lett. A}\ }\textbf {\bibinfo {volume} {230}},\ \bibinfo {pages} {105 }
  (\bibinfo {year} {1997})}\BibitemShut {NoStop}%
\bibitem [{\citenamefont {Ghosh}\ \emph {et~al.}(2016)\citenamefont {Ghosh},
  \citenamefont {Gupta}, \citenamefont {Jaswin},\ and\ \citenamefont
  {Nizami}}]{Ghosh:2016SM}%
  \BibitemOpen
  \bibfield  {author} {\bibinfo {author} {\bibfnamefont {S.}~\bibnamefont
  {Ghosh}}, \bibinfo {author} {\bibfnamefont {R.~K.}\ \bibnamefont {Gupta}},
  \bibinfo {author} {\bibfnamefont {K.}~\bibnamefont {Jaswin}}, \ and\ \bibinfo
  {author} {\bibfnamefont {A.~A.}\ \bibnamefont {Nizami}},\ }\href@noop {}
  {\bibfield  {journal} {\bibinfo  {journal} {JHEP}\ }\textbf {\bibinfo
  {volume} {2016}},\ \bibinfo {pages} {1 } (\bibinfo {year}
  {2016})}\BibitemShut {NoStop}%
\bibitem [{\citenamefont {Karkkainen}\ \emph {et~al.}(1994)\citenamefont
  {Karkkainen}, \citenamefont {Lacaze}, \citenamefont {Lacock},\ and\
  \citenamefont {B.Petersson}}]{Karkkainen:1994SM}%
  \BibitemOpen
  \bibfield  {author} {\bibinfo {author} {\bibfnamefont {L.}~\bibnamefont
  {Karkkainen}}, \bibinfo {author} {\bibfnamefont {R.}~\bibnamefont {Lacaze}},
  \bibinfo {author} {\bibfnamefont {P.}~\bibnamefont {Lacock}}, \ and\ \bibinfo
  {author} {\bibnamefont {B.Petersson}},\ }\href@noop {} {\bibfield  {journal}
  {\bibinfo  {journal} {Nucl. Phys. B}\ }\textbf {\bibinfo {volume} {415}},\
  \bibinfo {pages} {781} (\bibinfo {year} {1994})}\BibitemShut {NoStop}%
\bibitem [{\citenamefont {Nandkishore}\ \emph {et~al.}(2014)\citenamefont
  {Nandkishore}, \citenamefont {Huse},\ and\ \citenamefont
  {Sondhi}}]{Nandkishore:2014SM}%
  \BibitemOpen
  \bibfield  {author} {\bibinfo {author} {\bibfnamefont {R.}~\bibnamefont
  {Nandkishore}}, \bibinfo {author} {\bibfnamefont {D.~A.}\ \bibnamefont
  {Huse}}, \ and\ \bibinfo {author} {\bibfnamefont {S.~L.}\ \bibnamefont
  {Sondhi}},\ }\href@noop {} {\bibfield  {journal} {\bibinfo  {journal} {Phys.
  Rev. B}\ }\textbf {\bibinfo {volume} {89}},\ \bibinfo {pages} {245110}
  (\bibinfo {year} {2014})}\BibitemShut {NoStop}%
\bibitem [{\citenamefont {Falco}\ \emph {et~al.}(2009)\citenamefont {Falco},
  \citenamefont {Nattermann},\ and\ \citenamefont {Pokrovsky}}]{Falco:2009SM}%
  \BibitemOpen
  \bibfield  {author} {\bibinfo {author} {\bibfnamefont {G.~M.}\ \bibnamefont
  {Falco}}, \bibinfo {author} {\bibfnamefont {T.}~\bibnamefont {Nattermann}}, \
  and\ \bibinfo {author} {\bibfnamefont {V.~L.}\ \bibnamefont {Pokrovsky}},\
  }\href@noop {} {\bibfield  {journal} {\bibinfo  {journal} {Phys. Rev. B}\
  }\textbf {\bibinfo {volume} {80}},\ \bibinfo {pages} {104515} (\bibinfo
  {year} {2009})}\BibitemShut {NoStop}%
\bibitem [{\citenamefont {Falco}\ and\ \citenamefont
  {Fedorenko}(2015)}]{Falco:2015SM}%
  \BibitemOpen
  \bibfield  {author} {\bibinfo {author} {\bibfnamefont {G.~M.}\ \bibnamefont
  {Falco}}\ and\ \bibinfo {author} {\bibfnamefont {A.~A.}\ \bibnamefont
  {Fedorenko}},\ }\href {\doibase 10.1103/PhysRevA.92.023412} {\bibfield
  {journal} {\bibinfo  {journal} {Phys. Rev. A}\ }\textbf {\bibinfo {volume}
  {92}},\ \bibinfo {pages} {023412} (\bibinfo {year} {2015})}\BibitemShut
  {NoStop}%
\end{thebibliography}
\end{document}